\documentclass{emulateapj}

\usepackage{graphics}
\usepackage{epsfig}
\usepackage{natbib}
\shorttitle{UV Spectra of M dwarf Exoplanet Hosts}
\shortauthors{France et al.}
\begin{document}

\title{The Ultraviolet Radiation Environment Around M dwarf Exoplanet Host Stars\altaffilmark{*}}


\author{
Kevin France\altaffilmark{1}, 
Cynthia S. Froning\altaffilmark{1}, 
Jeffrey L. Linsky\altaffilmark{2},  
Aki Roberge\altaffilmark{3},
John T. Stocke\altaffilmark{1},
Feng Tian\altaffilmark{4},
Rachel Bushinsky\altaffilmark{1},
Jean-Michel D{\'e}sert\altaffilmark{5},
Pablo Mauas\altaffilmark{6},
Mariela Vieytes\altaffilmark{6},
Lucianne M. Walkowicz\altaffilmark{7}
}

\altaffiltext{*}{Based on observations made with the NASA/ESA $Hubble$~$Space$~$Telescope$, obtained from the data archive at the Space Telescope Science Institute. STScI is operated by the Association of Universities for Research in Astronomy, Inc. under NASA contract NAS 5-26555.}

\altaffiltext{1}{Center for Astrophysics and Space Astronomy, University of Colorado, 389 UCB, Boulder, CO 80309; kevin.france@colorado.edu}
\altaffiltext{2}{JILA, University of Colorado and NIST, 440 UCB, Boulder, CO 80309}
 \altaffiltext{3}{Exoplanets and Stellar Astrophysics Laboratory, 
 	NASA Goddard Space Flight Center, Greenbelt, MD 20771}	
\altaffiltext{4}{Center for Earth System Sciences, Tsinghua University, Beijing, China 100084}
\altaffiltext{5}{Division of Geological and Planetary Sciences, California Institute of Technology, Pasadena, CA 91125 }
\altaffiltext{6}{Instituto de Astronom�ay F�sica del Espacio (CONICET-UBA), C.C. 67 Sucursal 28, 1428, Buenos Aires, Argentina }
\altaffiltext{7}{Department of Astrophysical Sciences, Princeton University, Princeton, NJ 08544 }



\begin{abstract}

The spectral and temporal behavior of exoplanet host stars is a critical input to models of the chemistry and evolution of planetary atmospheres.  Ultraviolet photons influence the atmospheric temperature profiles and production of potential biomarkers on Earth-like planets around these stars.   At present, little observational or theoretical basis exists for understanding the ultraviolet spectra of M dwarfs, despite their critical importance to predicting and interpreting the spectra of potentially habitable planets as they are obtained in the coming decades.  Using observations from the {\it Hubble Space Telescope}, we present a study of the UV radiation fields around nearby M dwarf planet hosts that covers both FUV and NUV wavelengths. The combined FUV+NUV spectra are publically available in machine-readable format.  We find that all six exoplanet host stars in our sample (GJ 581, GJ 876, GJ 436, GJ 832, GJ 667C, and GJ 1214) exhibit some level of chromospheric and transition region UV emission.  No ``UV quiet'' M dwarfs are observed.  The bright stellar Ly$\alpha$ emission lines are reconstructed, and 
we find that the Ly$\alpha$ line fluxes comprise $\sim$~37~--~75\% of the total 1150~--~3100~\AA\ flux from most M dwarfs; $\gtrsim$~10$^{3}$ times the solar value.  
We develop an empirical scaling relation between Ly$\alpha$ and \ion{Mg}{2} emission,  
to be used when interstellar \ion{H}{1} attenuation precludes the direct observation of Ly$\alpha$.
The intrinsic unreddened flux ratio is $F$(Ly$\alpha$)/$F$(\ion{Mg}{2}) = 10~$\pm$~3.
The $F$(FUV)/$F$(NUV) flux ratio, a driver for abiotic production of the suggested biomarkers O$_{2}$ and O$_{3}$, is shown to be $\sim$~0.5~--~3 for all M dwarfs in our sample, $>$~10$^{3}$ times the solar ratio.  For the four stars with moderate signal-to-noise COS time-resolved spectra, we find UV emission line variability with amplitudes of 50~--~500\% on 10$^{2}$~--~10$^{3}$ s timescales.  This effect should be taken into account in future UV transiting planet studies, including searches for O$_{3}$ on Earth-like planets.  Finally, we observe relatively bright H$_{2}$ fluorescent emission from four of the M dwarf exoplanetary systems (GJ 581, GJ 876, GJ 436, and GJ 832).  Additional modeling work is needed to differentiate between a stellar photospheric or possible exoplanetary origin for the hot ($T$(H$_{2}$)~$\approx$~2000~--~4000 K) molecular gas observed in these objects.

\end{abstract}

\clearpage

\section{Introduction}

Stellar ultraviolet (UV) photons play a crucial role in the dynamics and chemistry of all types of planetary atmospheres.  Extreme-UV (EUV; 200~$\lesssim$~$\lambda$~$\lesssim$~911~\AA) photons capable of ionizing hydrogen heat the atmosphere and are the primary drivers of atmospheric escape in both the Solar System and exoplanetary systems. For short-period gas giant planets, EUV irradiation can heat the atmosphere to $\sim$~10$^{4}$ K~\citep{yelle04,murray-clay09}, resulting in atmospheric mass-loss~\citep{forcada11} and the inflated exospheres observed in \ion{H}{1} Ly$\alpha$~\citep{vidal03,lecavelier12} and metal absorption line~\citep{vidal04,linsky10,fossati10} transit spectra.  For terrestrial atmospheres, an increase in the EUV flux can elevate the temperature of the thermosphere by a factor of~$\gtrsim$~10~\citep{tian08}, potentially causing significant and rapid atmospheric mass-loss.   

Far-UV (FUV; 912~$\lesssim$~$\lambda$~$\lesssim$~1700~\AA) and near-UV (NUV; 1700~$\lesssim$~$\lambda$~$\lesssim$~4000~\AA) photons from the host star are the drivers of atmospheric chemistry through their effects on the photoexcitation and photodissociation rates of many abundant molecular species.  
Solar chromospheric emission (specifically \ion{H}{1} Ly$\beta$) excites H$_{2}$ fluorescence in gas giant planets, producing rovibrationally excited H$_{2}$ that catalyzes chemistry in the Jovian atmosphere~\citep{cravens87,kim94}.   In exo-Jovian atmospheres heated to $T$~$>$~1000K, Ly$\alpha$~fluorescence of both H$_{2}$ and CO may become an important excitation mechanism~\citep{wolven97}; this process may be an observable diagnostic for gas giant atmospheres outside the solar system~\citep{yelle04, france10a}.   H$_{2}$O, CH$_{4}$, and CO$_{2}$ are sensitive to FUV radiation, in particular the bright \ion{H}{1} Ly$\alpha$ line which has spectral coincidences with these species.  UV radiation can be important for the long-term habitability of Earth-like planets~\citep{buccino06}, and  
the combination of FUV and NUV photons can influence the molecular oxygen chemistry.  For example, the O$_{2}$ abundance can be significantly enhanced through the photodissociation of CO$_{2}$ and H$_{2}$O.  The subsequent formation of O$_{3}$ depends sensitively on the spectral and temporal behavior of the FUV and NUV radiation fields of the host star~\citep{tian12,goldman12}.   

The UV (FUV + NUV) spectra of solar-type stars have been studied extensively, both observationally and theoretically (e.g., Woods et al. 2009; Fontenla et al. 2011; Linsky et al. 2012c). \nocite{woods09,fontenla11,linsky12a}   By studying solar analogs of various ages, we also have an understanding of the evolution of G star spectra~\citep{ayres97,ribas05,linsky12a}.  By contrast, the observational and theoretical literature on the UV spectra of M dwarfs is sparse.  The photospheric UV continuum of M dwarfs is very low relative to solar-type stars due to their lower effective temperature. Bright chromospheric and transition region emission lines dominate the UV spectrum of M dwarfs, comprising a much larger fraction of the stellar luminosity than for solar-type stars.  Outside of a few well-studied flare stars (e.g., AU Mic, AD Leo, EV Lac, Proxima Cen), M dwarfs have largely been ignored by UV observers because most investigations were aimed at understanding the origin and nature of energetic events on low-mass stars.  There are very few UV observations of the older, quiescent M dwarfs that are most likely to harbor potentially habitable planets, and at present no theoretical atmosphere models exist that self-consistently predict the UV emission from an M dwarf.  

UV variability of M dwarf exoplanet host stars is also essentially unconstrained by observation at present.  M dwarfs display variability on many timescales and flare activity occurs with greater frequency and amplitude than on main-sequence G and K dwarfs~\citep{hawley96,west04,welsh07,walkowicz11}.  Because most flare activity is thought to be related to magnetic energy deposition in the corona, transition region, and chromosphere~\citep{haisch91}, UV flare activity has an amplitude comparable to or greater than that observed in optical light curves.   UV flare behavior on active M dwarfs is characterized by strong blue/NUV continuum emission and the enhancement of chromospheric and transition region emission lines~(e.g., Hawley \& Pettersen 1991; Hawley et al. 2003).  FUV flares may also be signpost for enhanced soft X-ray, EUV, and energetic particle arrival rates on planets orbiting these stars, all of which can play an important role for the chemistry and evolution of exoplanetary atmospheres~\citep{segura10} and the long-term habitability of these worlds~\citep{buccino07}.  

Given the paucity of existing observational data and lack of theoretical models of the UV spectra of M dwarfs, the most reliable incident radiation field for photochemical models of extrasolar planetary atmospheres is one created from direct observations of the host star.   Towards the goal of a comprehensive library of UV radiation fields for use as inputs in models of exoplanet atmospheres, we have carried out a pilot program to observe the spectral and temporal behavior of M dwarf exoplanet host stars, {\it Measurements of the Ultraviolet Spectral Characteristics of Low-mass Exoplanet host Stars} (MUSCLES).  

In this paper, we present observational results from the MUSCLES pilot program, spectrally and temporally resolved UV data for six M dwarf exoplanet host stars (GJ 581, GJ 876, GJ 436, GJ 832, GJ 667C, and GJ 1214) observed with  the {\it Hubble Space Telescope}-Cosmic Origins Spectrograph (COS) and Space Telescope Imaging Spectrograph (STIS).  We describe the targets and their planetary systems in Section 2.  Sections 3 and 4 describe the observations, the spectral reconstruction of the important stellar Ly$\alpha$ emission line, and the creation of light curves for chromospheric and transition region emission lines.  In Sections 5 and 6, we describe the spectral and temporal characteristics, respectively, of our target sample.  Section 7 places these results in context of models of terrestrial planetary atmospheres and discusses implications for UV transit studies.  We summarize our results in Section 8.  

\begin{figure}
\begin{center}
\epsfig{figure=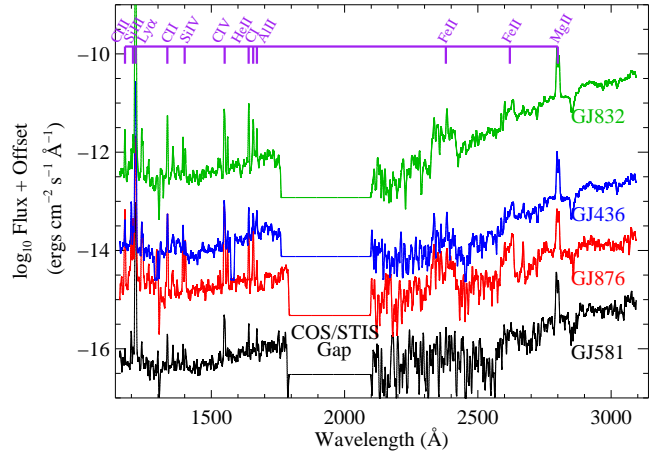,width=2.6in,angle=90}
\vspace{+0.0in}
\caption{
\label{cosovly} The observed 1150~--~3140~\AA\ fluxes from the M dwarf exoplanet host stars in the MUSCLES sample with complete (Ly$\alpha$ + FUV + NUV) datasets.  
The 1760~--~2100~\AA\ emission has been omitted because the flux levels are below the STIS G230L detection level. 
For display purposes, the data have been convolved with a 3~\AA\ FWHM Gaussian kernel and offset as follows: GJ 581, no offset; GJ 876, log$_{10}$F$_{\lambda}$ + 1.2; GJ 436, log$_{10}$F$_{\lambda}$ + 2.4; GJ 832, log$_{10}$F$_{\lambda}$ + 3.6. 
 }
\end{center}
\end{figure}

\begin{figure*}[t]
\begin{center}
\epsfig{figure=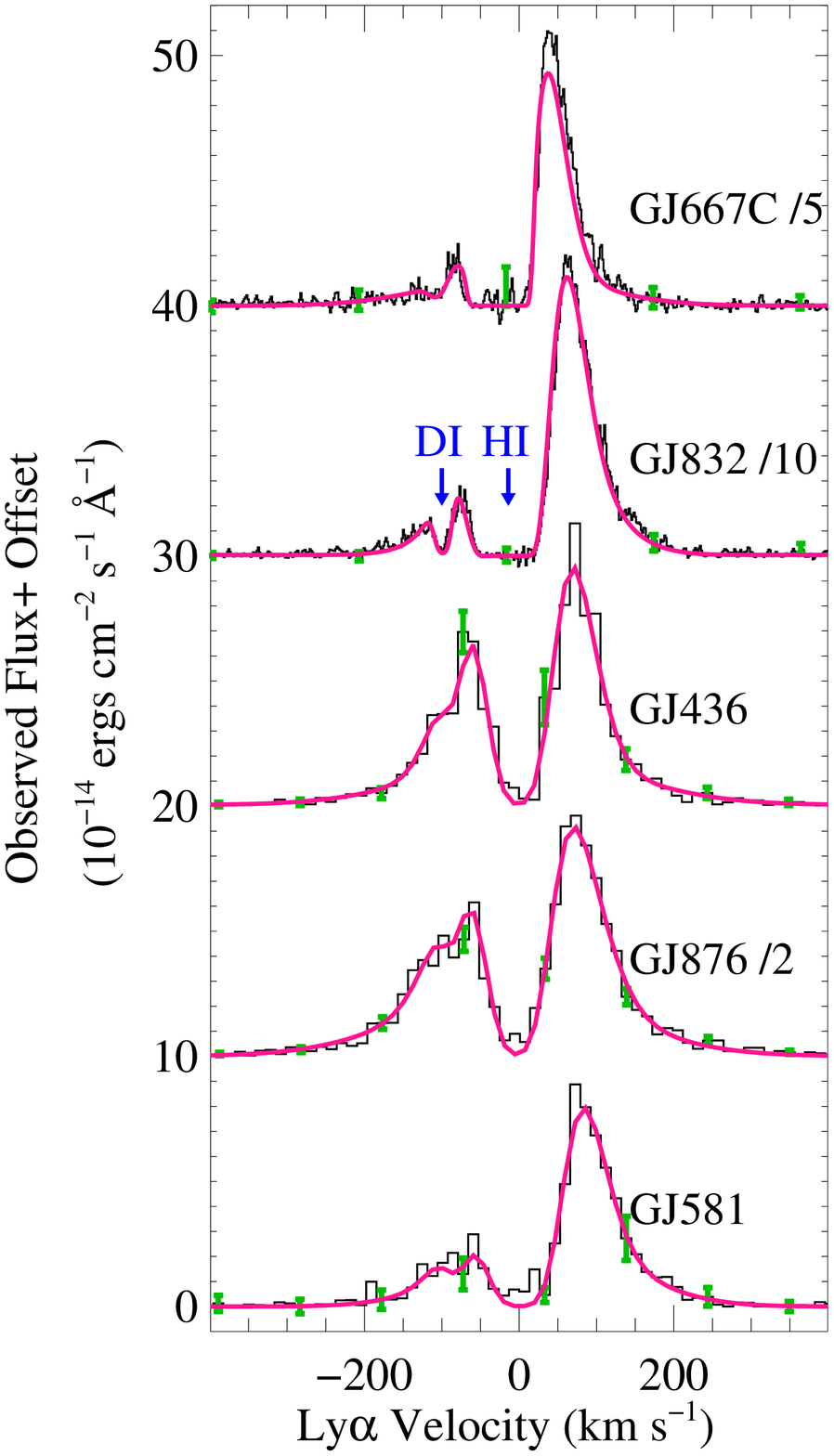,width=3.0in,angle=00}
\epsfig{figure=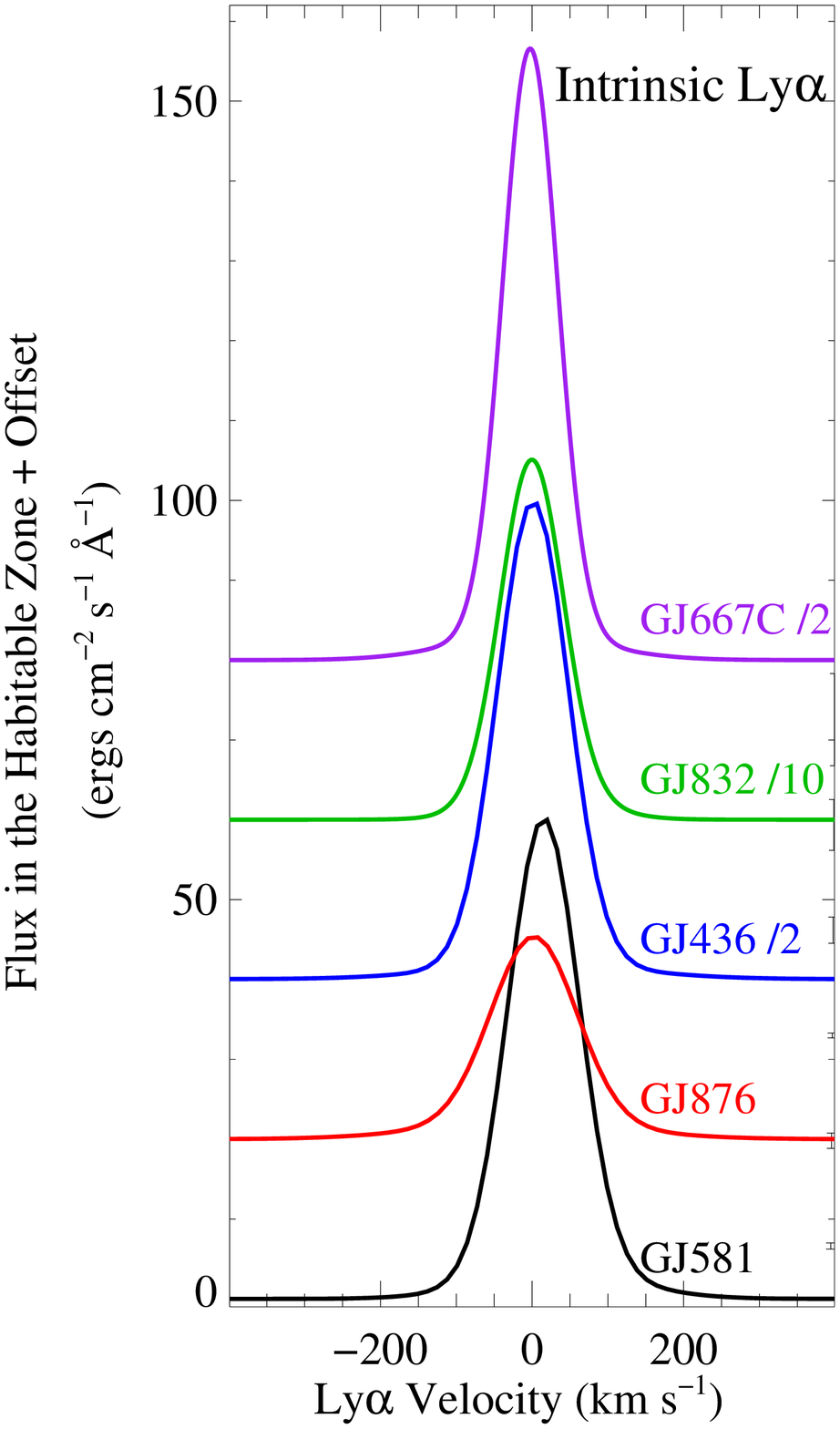,width=3.0in,angle=00}
\vspace{+0.0in}
\caption{
\label{cosovly} ($left$) $HST$-STIS E140M and G140M Ly$\alpha$ profiles ({\it black histograms}).  The two-component Gaussian emission line + interstellar absorption model is overplotted in pink.  The neutral hydrogen and deuterium absorption components are marked with blue arrows.  The spectra are scaled as marked and offset in +10~$\times$~10$^{-14}$ erg cm$^{-2}$ s$^{-1}$ \AA$^{-1}$ steps, photometric uncertainties are displayed as green error bars.   ($right$) Intrinsic two-component Gaussian model to the stellar Ly$\alpha$ emission line, scaled to the flux at 0.16~AU, in the habitable zone of these M dwarf exoplanet host stars.  The spectra are scaled as marked and offset in +20 erg cm$^{-2}$ s$^{-1}$ \AA$^{-1}$ steps. 
 }
\end{center}
\end{figure*}
\section{Targets}

In this section, we briefly summarize the stellar and planetary system properties of our M dwarf targets.  Optical spectra of most of our targets have been published in the literature, and while we do not make a detailed comparison of the UV and visible emission lines here, it is valuable to place the MUSCLES targets in the context of traditional optical activity indicators.  The stars in our sample would traditionally be considered ``optically inactive'', based on their H$\alpha$ absorption spectra~\citep{gizis02}.  However, all of our stars with measured \ion{Ca}{2} H and K  profiles show weak but detectable emission (equivalent widths, EW(\ion{Ca}{2}) $>$~0), indicating that at least a low level of chromospheric activity is present in these stars~\citep{rauscher06,walkowicz09}.  Our stars show H$\alpha$ in absorption with equivalent widths in the range $-$0.4~$<$~EW(H$\alpha$)~$<$~$-$0.2~\AA\ and \ion{Ca}{2} emission with equivalent widths in the range 0.2~$<$~EW(\ion{Ca}{2})~$<$~0.5~\AA.  Adopting the M dwarf classification from~\citet{walkowicz09}, these stars would be referred to as possessing intermediate chromospheres, or weakly active M dwarfs.

\begin{figure}
\begin{center}
\epsfig{figure=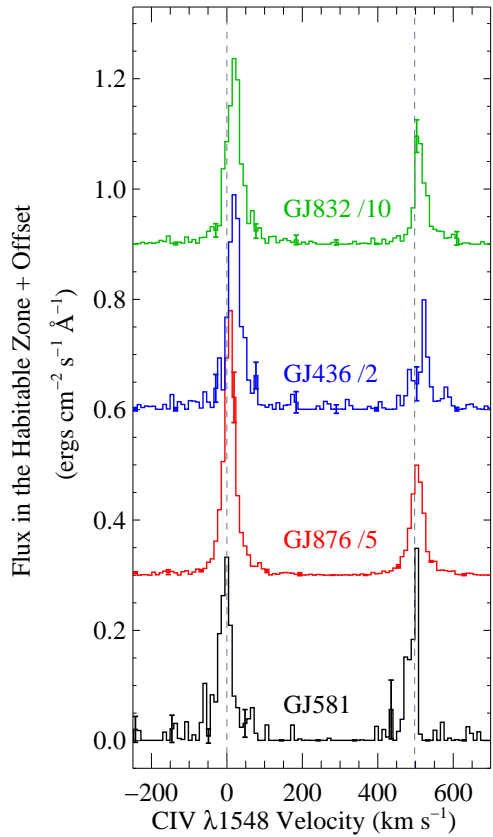,width=2.8in,angle=00}
\vspace{+0.0in}
\caption{ \ion{C}{4} $\lambda$$\lambda$1548,1550~\AA\ resonance doublets for the four stars with Ly$\alpha$ detections and FUV spectra, plotted in the heliocentric velocity frame.  The emission lines have been scaled to the flux at 0.16~AU, in the habitable zone of these M dwarf exoplanet host stars, and offset in 0.3 erg cm$^{-2}$ s$^{-1}$ \AA$^{-1}$ steps. The dashed lines are the rest velocities of the 1548~\AA\ and 1550~\AA\ lines, respectively.  
\label{cosovly} 
 }
\end{center}
\end{figure}
\begin{deluxetable}{lcccc}
\tabletypesize{\normalsize}
\tablecaption{MUSCLES $HST$ Observations. \label{lya_lines}}
\tablewidth{0pt}
\tablehead{
\colhead{Target}  &  \colhead{Date} & \colhead{Instrument, Mode}  & \colhead{PID}    &  \colhead{$T_{exp}$ (s) }  }
\startdata
GJ 581 & 2012 Apr 18 & STIS, G140M  & 12034 & 1107\\
GJ 581 & 2012 Apr 18 & STIS, G230L  & 12034  & 265 \\
GJ 581 & 2011 Jul 20 & COS, G130M  & 12034  & 1478 \\
GJ 581 & 2011 Jul 20 & COS, G160M  & 12034  & 900 \\
\tableline
GJ 876 & 2011 Nov 12 & STIS, G140M  & 12464  & 1138 \\
GJ 876 & 2011 Nov 12 & STIS, G230L  & 12464  & 267 \\
GJ 876 & 2012 Jan 05 & COS, G130M  & 12464  & 2017 \\
GJ 876 & 2012 Jan 05 & COS, G160M  & 12464  & 2779 \\
\tableline
GJ 436 & 2010 Jan 05 & STIS, G140M  & 11817  & 1762 \\
GJ 436 & 2012 May 10 & STIS, G230L  & 12464  & 1665 \\
GJ 436 & 2012 Jun 23 & COS, G130M  & 12464  & 3372 \\
GJ 436 & 2012 Jun 23 & COS, G160M  & 12464  & 4413 \\
\tableline
GJ 832 & 2011 Jun 09 & STIS, E140M  & 12035  & 2572 \\
GJ 832 & 2012 Jun 10 & STIS, E230H  & 12035  & 3135 \\
GJ 832 & 2012 Apr 10 & STIS, G230L  & 12464  & 917 \\
GJ 832 & 2012 Jul 28  & COS, G130M  & 12464  & 2163 \\
GJ 832 & 2012 Jul 28 & COS, G160M  & 12464  & 2925 \\
\tableline
GJ 667C & 2011 Sept 04 & STIS, E140M  & 12035  & 2396 \\
GJ 667C & 2011 Sept 04 & STIS, E230H  & 12035  & 3023 \\
\tableline
GJ 1214 & 2011 Apr 27 & STIS, G140M  & 12165  & 7282 \\
GJ 1214 & 2012 Aug 12 & STIS, G230L  & 12464  & 1620 \\
GJ 1214 & 2012 Aug 04 & COS, G130M  & 12464  & 3289 \\
GJ 1214 & 2012 Aug 04 & COS, G160M  & 12464  & 4368 
\enddata
\end{deluxetable}
{\it GJ 581}~--~ GJ 581 is an M2.5 dwarf at a distance of 6.3 pc.  It is estimated to have an age of 8~$\pm$~1 Gyr~\citep{selsis07} and a somewhat subsolar metallicity, [Fe/H]~=~$-$0.10~--~$-$0.02~\citep{johnson09,rojas10}.  GJ 581 is not detected in X-ray surveys (log$_{10}$$L_{X}$~$<$~26.89 erg s$^{-1}$; Poppenhaeger et al. 2010) and its optical spectrum displays H$\alpha$ in absorption, therefore chromospheric and coronal activity are thought to be low for this target.\nocite{poppenhaeger10}  GJ 581 has one of the richest known planetary systems, with possibly up to six planets (four confirmed) including several of Earth/super-Earth mass~\citep{mayor09,tuomi11}.  GJ 581d is a super-Earth ($M_{P}$~$\approx$~6~$M_{\oplus}$) that resides on the outer edge of the habitable zone (HZ; $a_{P}$~=~0.22 AU; Wordsworth et al. 2011; von Braun et al. 2011). \nocite{wordsworth11,vonbraun11}  

{\it GJ 876}~--~ GJ 876 is an M4 dwarf at 4.7 pc, and is the only planet-hosting M dwarf with a well-characterized UV spectrum prior to the present work~\citep{walkowicz08,france12b}.  GJ 876 has super-solar metallicity ([Fe/H] = 0.37~--~0.43; Johnson \& Apps 2009; Rojas-Ayala et al. 2010), and differing estimates on the stellar rotation period (40~$\leq$~$P_{*}$~$\lesssim$~97 days) result in large uncertainties in the age estimate for this system, 0.1~--~5 Gyr~\citep{rivera05,rivera10,correia10}.\nocite{johnson09,rojas10}  While the star would be characterized as weakly active based on its H$\alpha$ absorption spectrum, UV and X-ray observations have shown the presence of an active upper atmosphere (Walkowicz et al. 2008; France et al. 2012a; log$_{10}$$L_{X}$~$=$~26.48 erg s$^{-1}$; Poppenhaeger et al. 2010).  GJ 876 has a rich planetary system, with four planets ranging from a super-Earth (GJ 876d, $M_{P}$~$\approx$~6.6~$M_{\oplus}$) in a short-period orbit ($a_{P}$~=~0.02 AU; Rivera et al. 2010) to two Jovian-mass planets in the HZ (GJ 876b, $M_{P}$~$\approx$~2.27~$M_{Jup}$, $a_{P}$~=~0.21 AU; GJ 876c, $M_{P}$~$\approx$~0.72~$M_{Jup}$, $a_{P}$~=~0.13 AU; Rivera et al. 2010).

\begin{deluxetable*}{lccccc}
\tabletypesize{\normalsize}
\tablecaption{Stellar Ly$\alpha$ Emission Line Widths and Interstellar Atomic Hydrogen\tablenotemark{a}. \label{lya_lines}}
\tablewidth{0pt}
\tablehead{
\colhead{Target} & \colhead{$F_{narrow}$/$F_{broad}$} & \colhead{FWHM$_{narrow}$} & \colhead{FWHM$_{broad}$}   &  \colhead{log$_{10}$ $N$(HI)} & \colhead{$b_{HI}$}    \\
\colhead{} &   & \colhead{(km s$^{-1}$)} &  \colhead{(km s$^{-1}$)} & \colhead{(cm$^{-2}$) }  & \colhead{(km s$^{-1}$)}
}
\startdata
GJ 581\tablenotemark{b}  & 8.4  & 110 & 228  & 18.35~$\pm$~0.06  &  10.1  \\
GJ 876\tablenotemark{b}   &  5.0 & 132 & 303  & 18.06~$\pm$~0.03  &  8.0   \\
GJ 436\tablenotemark{b}  &   11.8 & 109 & 294  & 18.19~$\pm$~0.04  &  7.4   \\

GJ 832\tablenotemark{c}  &   7.6 & 96 & 163  & 18.47~$\pm$~0.02  &  9.7   \\
GJ 667C\tablenotemark{c} &   8.9 & 85 & 233  & 18.07~$\pm$~0.03  &  12.7   \\
\tableline\tableline
AD Leo\tablenotemark{c}       &  2.9  & 166 & 412  & 18.47~$\pm$~0.01  &  9.0   \\
HD 189733\tablenotemark{b,d} &  3.5 &  170 & 480  & 18.45  &  10
\enddata
\tablenotetext{a}{- Ly$\alpha$ profiles reconstructed using the iterative least-squares technique described in \S4.1. } 
\tablenotetext{b}{STIS G140M observations.} 
\tablenotetext{c}{STIS E140M observations.}  
\tablenotetext{d}{HD 189733 reconstruction failed to converge,  the presented profile parameters provide a fit to the observed line profile that is qualitatively similar to the formal fits for the M dwarf spectra.  } 
\end{deluxetable*}

{\it GJ 436}~--~ GJ 436 is an M3 dwarf star located at a distance of 10.3 pc.  It has a 45 day rotation period, a relatively old age ($\sim$~6$^{+4}_{-5}$~Gyr; Torres 2007), and may have a super-solar metallicity ([Fe/H]~=~0.00~--~0.25; Johnson \& Apps 2009; Rojas-Ayala et al. 2010).  GJ 436 does show signs of an active corona~(log$_{10}$$L_{X}$~$=$~27.16 erg s$^{-1}$; Poppenhaeger et al. 2010), and its chromospheric Ly$\alpha$ emission has been observed by~\citet{ehrenreich11}.  GJ 436 is notable for its well-studied transiting Neptune mass planet~\citep{butler04,pont09}, orbiting at a semi-major axis of ~$\approx$~0.03 AU, interior to its HZ (0.16~--~0.31~AU; von Braun et al. 2012).\nocite{vonbraun12,torres07}   Additional low-mass planets may also be present in this system~\citep{stevenson12}.  

{\it GJ 832}~--~ GJ 832 is an M1 dwarf at $d$~=~4.9 pc.  GJ 832 is not as well characterized as other targets in our sample; an age determination for this star is not available.  Coronal X-rays have been detected from GJ 832, log$_{10}$$L_{X}$~$=$~26.77 erg s$^{-1}$ (Poppenhaeger et al. 2010).  This subsolar metallicity star ([Fe/H] = $-$0.12; Johnson \& Apps 2009) hosts a 0.64~$M_{Jup}$ mass planet in a 9.4 year orbit ($a_{P}$~=~3.4 AU; Bailey et al. 2009).\nocite{bailey09}

{\it GJ 667C}~--~ GJ 667C (M1.5V) is a member of a triple star system (GJ 667AB is a K3V + K5V binary) at a distance of 6.9 pc.  This 2~--~10 Gyr M dwarf~\citep{anglada_escude12} is metal-poor ([Fe/H] = $-$0.59~$\pm$~0.10, based on an analysis of GJ 667AB; Perrin et al. 1988) and may host as many as three planets, including a super-Earth mass planet (GJ 667Cc, $M_{P}$~$\approx$~4.5~$M_{\oplus}$, $a_{P}$~=~0.12 AU) orbiting in the HZ (0.11~--~0.23 AU; Anglada-Escud{\'e} et al. 2012).\nocite{perrin88}

{\it GJ 1214}~--~ GJ 1214 is a late M dwarf (M6V) at 13 pc, making it the coolest and most distant of the targets in the MUSCLES pilot study.  It has an age of~6~$\pm$~3 Gyr~\citep{charbonneau09}, a super-solar metallicity ([Fe/H]~=~+0.39~$\pm$~0.15; Berta et al. 2011; Rojas-Ayala et al. 2010), and shows signs of optical flare activity~\citep{kundurthy11}.  GJ 1214b is a transiting super-Earth ($M_{P}$~$\approx$~6.5~$M_{\oplus}$, $a_{P}$~=~0.014 AU; Charbonneau et al. 2009), possibly harboring a dense, water-rich atmosphere~\citep{bean10,desert11}.

\section{Observations}

The MUSCLES observing plan uses the two primary ultraviolet spectrographs on $HST$ to create quasi-continuous M dwarf exoplanet host star data from 1150~--~3140~\AA.  Data for the MUSCLES program were obtained as part of $HST$ GTO and GO programs 12034, 12035, and 12464, acquired between 2011 June and 2012 August.  COS has a factor of $>$ 10$\times$ more effective area than the medium resolution modes of STIS in the FUV, and its low detector background and grating scatter make it factors of $\sim$~50$\times$ more efficient for the study of faint far-UV chromospheric and transition region emission from M dwarfs~\citep{green12}.  We used COS, in its time-tag (TTAG) acquisition mode, to observe the FUV spectra of our target stars and study time variability on scales from minutes to hours (\S4.2 and \S6).  Multiple central wavelengths and focal-plane offset (FP-POSs) settings with the COS G130M and G160M modes 
were used to create a continuous FUV spectrum from 1145~--~1795~\AA.  These modes provide a point-source resolution of $\Delta$$v$~$\approx$~17 km s$^{-1}$ with 7 pixels per resolution element~\citep{osterman11}.  

Because COS is a slitless spectrograph, observations at \ion{H}{1} Ly$\alpha$ are heavily contaminated by geocoronal emission.  We therefore observed the stellar Ly$\alpha$ profile with STIS, using either the G140M/cenwave 1222 mode ($\Delta$$v$~$\approx$~30 km s$^{-1}$) through the 52\arcsec~$\times$~0.1\arcsec\ slit, or the E140M mode ($\Delta$$v$~$\approx$~7.5 km s$^{-1}$) through the 0.2\arcsec~$\times$~0.2\arcsec\ slit for the brightest targets.   In order to include the important Ly$\alpha$ emission line for GJ 436 and GJ 1214, we downloaded STIS G140M observations of these stars from the MAST archive (program IDs 11817 and 12165, respectively).  The Ly$\alpha$ profile of GJ 436 has been described in~\citet{ehrenreich11}.     In order to maximize the combination of wavelength coverage and sensitivity, near-UV spectra were observed with the STIS G230L/cenwave 2376 mode ($\Delta$$v$~$\approx$~600 km s$^{-1}$) through the 52\arcsec~$\times$~0.1\arcsec\ slit.  In order to constrain the effects of interstellar absorption on the stellar \ion{Mg}{2} emission profile, we also acquired \ion{Mg}{2} spectra with the STIS E230H mode ($\Delta$$v$~$\approx$~2.6 km s$^{-1}$) through the 0.2\arcsec~$\times$~0.2\arcsec\ slit for GJ 832 and GJ 667C.  

Exposure times were generally short due to the pilot-study nature of this project; considerably longer integrations will be essential for future observations to measure UV flare frequencies in exoplanet host stars and to produce high quality line profiles of important chromospheric and transition region tracers such as \ion{C}{2} $\lambda$1335, \ion{N}{5} $\lambda$1240, and \ion{Si}{4} $\lambda$1400.   A complete list of the M dwarf observations used in this work is presented in Table 1, and complete spectra for the four targets in which Ly$\alpha$, FUV, and NUV emission is observed are displayed in Figure 1.  

\begin{figure}
\begin{center}
\epsfig{figure=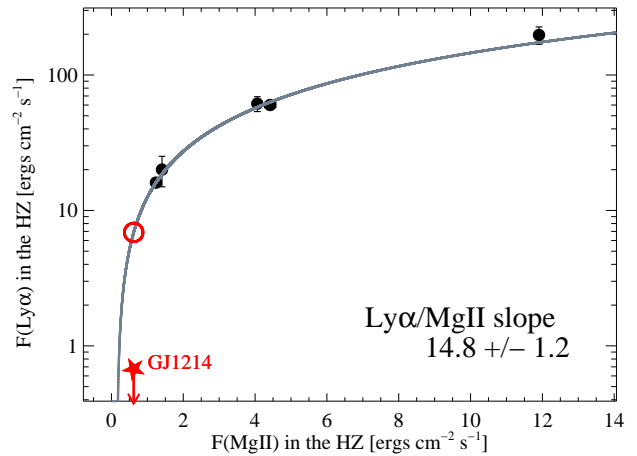,width=2.6in,angle=90}
\vspace{+0.0in}
\caption{ Intrinsic, reconstructed Ly$\alpha$ emission fluxes compared with the observed \ion{Mg}{2} ($h$ + $k$) flux, both scaled to the flux at 0.16 AU from the star.  The Ly$\alpha$ and \ion{Mg}{2} fluxes are observed to be well correlated.  
The slope of the relation is 14.8~$\pm$~1.2.  After correcting for the effects of interstellar absorption of the \ion{Mg}{2} lines, the $F$(Ly$\alpha$)/$F$(\ion{Mg}{2}) ratio for weakly active M dwarf exoplanet host stars is 10~$\pm$~3.
The expected Ly$\alpha$ emission line strength of GJ 1214 is shown as the open red circle.  The observed GJ 1214 Ly$\alpha$ emission line flux upper limit is shown as the filled red star.  
\label{cosovly} 
 }
\end{center}
\end{figure}

\begin{figure}
\begin{center}
\epsfig{figure=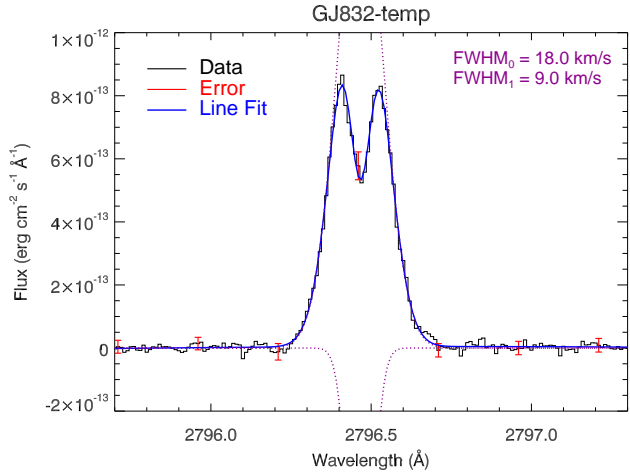,width=2.6in,angle=90}
\vspace{-0.1in}
\caption{
\label{cosovly} STIS E230H profile of the \ion{Mg}{2} $k$ line ($\lambda_{o}$~=~2795.528~\AA) of GJ 832.  The dotted magenta lines are the stellar emission and interstellar absorption components of the fit, respectively.   
Fitting the high-resolution echelle spectra of the \ion{Mg}{2} $k$ and $h$ lines of GJ 832 and GJ 667 C, we estimate that the interstellar correction for the \ion{Mg}{2} flux is ~$\approx$~30~--~35\%.  
 }
\end{center}
\end{figure}

In order to place the M dwarf exoplanet host stars in context with other well-studied cool stars, we assembled archival spectra of AD Leo (M3.5Ve), HD 189733 (K1V), and the Sun (G2V).  AD Leo is one of the best-studied flare stars, and despite its extreme activity levels is has been the object of choice for models of extrasolar planets orbiting M dwarfs, mainly because of the lack of reasonable alternatives.  We created a complete UV spectrum of AD Leo by combining STIS E140M monitoring observations~\citep{hawley03} with an ``average'' NUV spectrum from $IUE$.  The $IUE$ spectrum is the average of flare and quiescent states, scaled to the FUV flux from STIS.   HD189733 is an active K dwarf hosting the prototypical transiting hot Jupiter.  While high-quality COS spectra of HD 189733 exist in the archive (e.g., Linsky et al. 2012; Haswell et al. 2012), we elected to use the combined NUV+FUV (STIS E140M + E230M) spectra of a surrogate star taken from the STARCAT library~\citep{ayres10} to eliminate uncertainties about temporal variability between different observations.  We chose HD37394 (K1V variable star; $V$~=~6.23) because it is a good match in stellar mass and activity level to HD 189733, and because the FUV spectra of the two stars are qualitatively similar.  The UV spectrum of HD37394 was then scaled to the V-magnitude of HD 189733 ($V$~=~7.77).  The spectrum of the quiet Sun was taken from~\citet{woods09}, and serves as the prototype main-sequence G-type star.  

\section{Intrinsic Ly$\alpha$ Profile Reconstruction and Time Variability Analysis}

\subsection{Ly$\alpha$ Reconstruction} 

The stellar Ly$\alpha$ emission line dominates the UV output of M dwarfs, containing approximately as much energy as the rest of the FUV+NUV spectrum combined~\citep{france12b}.  Therefore, measurements of Ly$\alpha$ emission provide an important constraint on the source term for calculations of the heating and photochemistry of exoplanetary atmospheres~\citep{linsky12c}.  Due to resonant scattering of neutral hydrogen and deuterium in the interstellar medium (ISM), the intrinsic Ly$\alpha$ radiation field cannot be directly measured, even in the nearest stars (log$_{10}$ $N$(HI)~$\lesssim$~18.5).  In order to produce an accurate estimate of the local Ly$\alpha$ flux incident on the planets around our target stars, the Ly$\alpha$ profiles observed by STIS must be reconstituted.  We do this using a technique that simultaneously fits the observed Ly$\alpha$ emission line profile and the interstellar medium component, using the MPFIT routine to minimize $\chi^{2}$ between the fit and data~\citep{markwardt09}.

The intrinsic Ly$\alpha$ emission line is approximated as a two-component Gaussian, as suggested for transition region emission lines from late-type stars by~\citet{wood97}.  This approximation is further justified by the reconstructed Ly$\alpha$ profiles for M dwarfs, which show approximately Gaussian line shapes and little evidence for self-reversal compared to F, G, and K dwarfs~\citep{wood05}.     
The Ly$\alpha$ emission components are characterized by an amplitude, FWHM, and velocity centroid.
Interstellar absorption is parameterized by the column density of neutral hydrogen ($N$(HI)), Doppler $b$ parameter, D/H ratio, and the velocity of the interstellar absorbers.  
We assumed a fixed D/H ratio (D/H = 1.5~$\times$~10$^{-5}$; Linsky et al. 1995; 2006).\nocite{linsky95,linsky06}    Degeneracy between the two emission components makes errors difficult to define for the individual fit parameters.
We estimate the uncertainty on the integrated Ly$\alpha$ fluxes from our technique by comparing the reconstructed Ly$\alpha$ flux for a range of initial guesses on the parameters.   The uncertainty on the integrated intrinsic flux is $\approx$~10~--~20~\% for the stars observed with E140M (GJ 832 and GJ 667C) and $\approx$~15~--~30~\% for the stars observed with G140M (GJ 581, GJ 876, GJ 436).  In order to test our methodology, we fitted the STIS E140M Ly$\alpha$ profile of AU Mic. Our total integrated flux agrees with that presented in~\citet{wood05} to $\approx$~5\%, with an identical derivation of the interstellar atomic hydrogen column on the line of sight (log$_{10}$ $N$(HI)~=~18.36~$\pm$~0.01).     

Employing our iterative least-squares reconstruction technique, we also derived the intrinsic Ly$\alpha$ spectrum of HD 189733. Using the 2010 April 06 STIS G140M spectra~\citep{lecavelier12}, we find an intrinsic flux of
 $F$(Ly$\alpha$)~=~7.5~$\times$~10$^{-13}$ erg cm$^{-2}$ s$^{-1}$ and an interstellar hydrogen column density log$_{10}$ $N$(HI)~=~18.45.   We also refitted the intrinsic Ly$\alpha$ emission from AD Leo, finding $F$(Ly$\alpha$)~=~7.5~$\times$~10$^{-12}$ erg cm$^{-2}$ s$^{-1}$ and an interstellar hydrogen column density log$_{10}$ $N$(HI)~=~18.47~$\pm$~0.02.   This value agrees with the Ly$\alpha$ reconstruction from~\citet{wood05} to $\approx$~30\%, which gives us confidence in the iterative technique, even for sightlines with multiple interstellar velocity components.  

\begin{figure}[t]
\begin{center}
\epsfig{figure=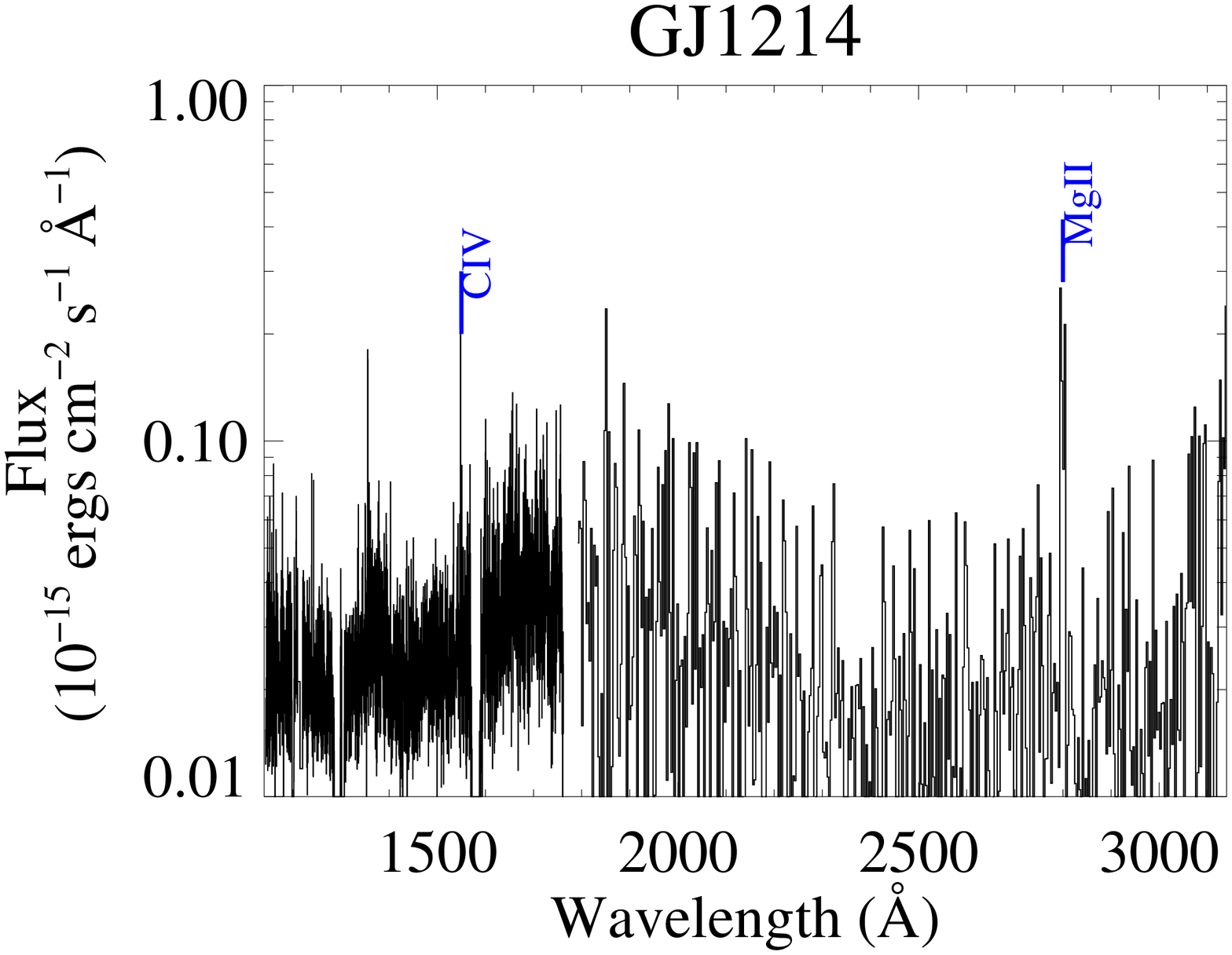,width=2.4in,angle=90}
\epsfig{figure=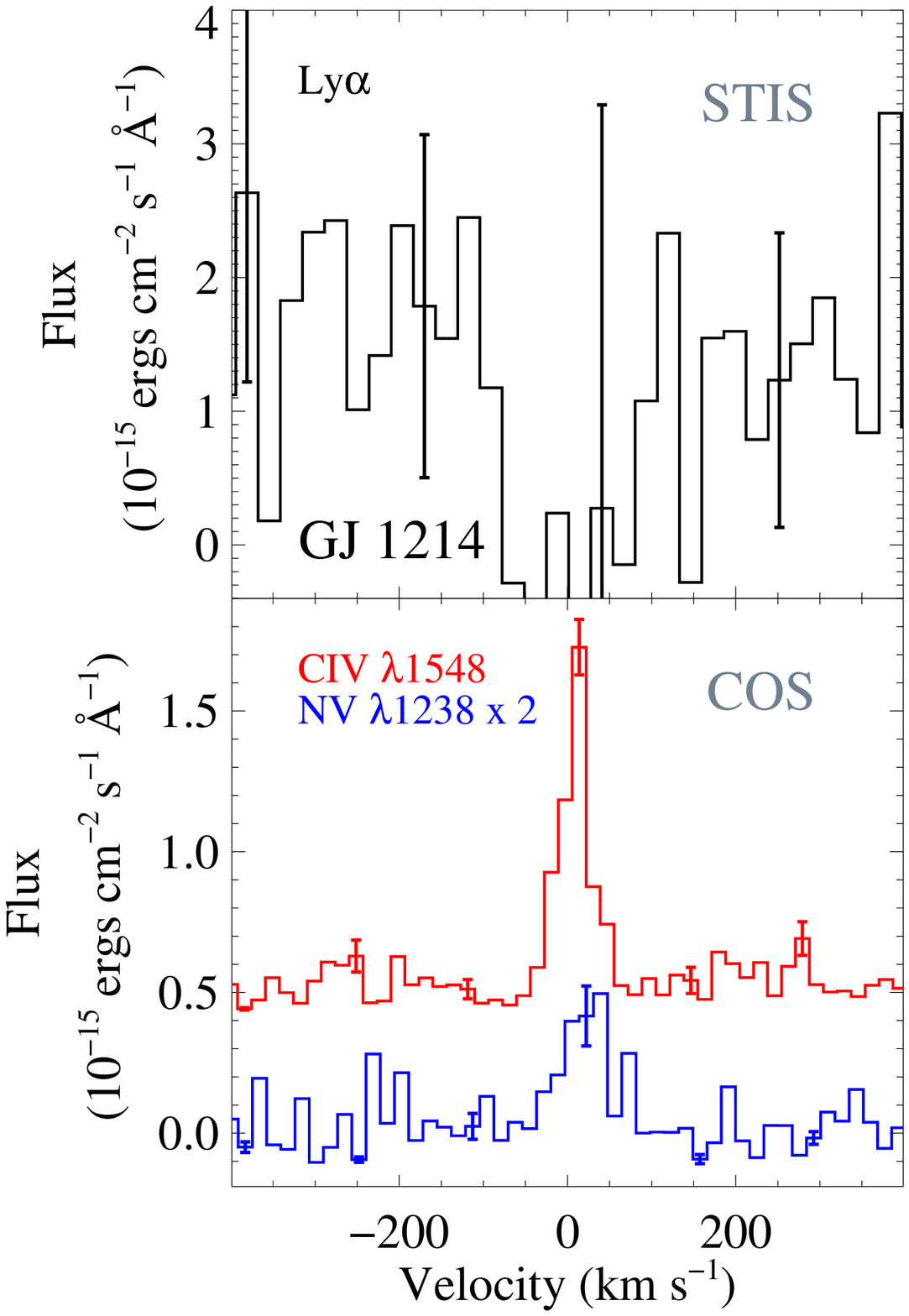,width=2.8in,angle=00}
\vspace{+0.1in}
\caption{ ($top$) $HST$-COS + STIS 1140~--~3140~\AA\ spectrum of GJ 1214.  Emissions from \ion{C}{4} and \ion{Mg}{2} are the only strongly detected features.  At bottom, we show the $\pm$~400 km s$^{-1}$ spectrum around \ion{H}{1} Ly$\alpha$ ({\it upper panel, black histogram}), \ion{C}{4} $\lambda$1548 ({\it lower panel, red histogram}), and \ion{N}{5} $\lambda$1239 ({\it lower panel, blue histogram}) scaled up by a factor of 2 for display.  The \ion{Mg}{2} and \ion{C}{4} lines demonstrate the presence of an active atmosphere; several possibilities for the non-detection of Ly$\alpha$ are presented in \S5.1.1.  
\label{cosovly} 
 }
\end{center}
\end{figure}

We present the individual Ly$\alpha$ spectra in Figure 2, and narrow/broad emission component ratios ($F_{narrow}$/$F_{broad}$), spectral line widths, \ion{H}{1} column densities, and Doppler $b$-values in Table 2.  We find $F_{narrow}$/$F_{broad}$ to be in the range 5~--~12 for the MUSCLES stars, again independent of the resolution of the observations on which the reconstructions are based.  For comparison, we find $F_{narrow}$/$F_{broad}$~$\sim$~3 for the more active M dwarf (AD Leo) and more massive star (HD 189733).  
Reconstructed Ly$\alpha$ fluxes are given in Tables 3, 4, and 5.  GJ 1214 has no Ly$\alpha$ emission detected, and we discuss this in greater detail in \S5.1.1.

\subsection{Timing Analysis} 

In our initial study of GJ 876, we described the large chromospheric/transition region flare observed in several UV emission lines during our COS observations~\citep{france12b}.  We have carried out a similar analysis on the entire MUSCLES sample to constrain the heretofore unknown level of UV variability in optically quiet M dwarf exoplanet host stars.  Light curves in several chromospheric and transition region lines were extracted from the calibrated three-dimensional data by exploiting the time-tag capability of the COS microchannel plate detector~\citep{france10b}.  We extract a [$\lambda_{i}$,$y_{i}$,$t_{i}$] photon list (where $\lambda$ is the wavelength of the photon, $y$ is the cross-dispersion location, and $t$ is the photon arrival time) from each exposure $i$ and combine these to create a master [$\lambda$,$y$,$t$] photon list. The total number of counts in a [$\Delta$$\lambda$,$\Delta$$y$] box is integrated over a timestep $\Delta$$t$.   We use a flux-dependent timestep of $\Delta$$t$ = 40~--~120s for the MUSCLES targets.  The instrument background level is computed in a similar manner, with the background integrated over the same wavelength interval as the emission lines, but physically offset below the science region in the cross-dispersion direction.  The S/N was not high enough to temporally resolve the GJ 1214 data.  

\section{Stellar UV Emission Lines and the FUV/NUV Ratio} 

\subsection{Chromospheric and Transition Region Emission}

Chromospheric and transition region emission lines are observed in all of the MUSCLES spectra, suggesting that many M dwarf exoplanet host stars have UV-active atmospheres (Figure 1).  The chromosphere is observed in \ion{H}{1} Ly$\alpha$, \ion{C}{2} $\lambda$$\lambda$1334,1335, \ion{Al}{2} $\lambda$1671, \ion{Fe}{2} multiplets near 2400 and 2600~\AA, and \ion{Mg}{2}~$\lambda$$\lambda$2796,2803; lines that are formed from $\sim$~1~--~3~$\times$~10$^{4}$ K.  The transition region is traced by lines with formation temperatures from $\sim$~0.6~--~1.6~$\times$~10$^{5}$ K, including the \ion{C}{3} 1175 multiplet, \ion{Si}{3} $\lambda$1206, \ion{O}{5} $\lambda$1218, \ion{N}{5} $\lambda$$\lambda$1239,1243, \ion{Si}{4} $\lambda$$\lambda$1394,1403, \ion{C}{4} $\lambda$$\lambda$1548,1550, and \ion{He}{2} $\lambda$1640.  \ion{C}{2} and \ion{C}{4} line fluxes for GJ 581, GJ 876, GJ 436, and GJ 832 have been published in~\citet{linsky12c}, and a comprehensive set of stellar emission line measurements will be presented in a future work.  In Table 5, we present the reconstructed Ly$\alpha$ fluxes and the observed fluxes of \ion{C}{4} and \ion{Mg}{2}. 

Figure 3 displays the \ion{C}{4} profiles for GJ 581, GJ 876, GJ 436, and GJ 832 scaled to the flux level at 0.16 AU from the parent star, a value near the center of the HZ for the MUSCLES targets\footnote{~\citet{vonbraun12} compute the HZ of GJ 436 to extend 0.16~--~0.31 AU, so 0.16 AU may represent the inner edge of the HZ for this system.}.   While we conclude that most M dwarf exoplanet hosts have a non-zero local UV radiation field, the amplitude of this field is quite variable.  The HZ \ion{C}{4} fluxes from these four stars varies by a factor of 10 from the weakest (GJ 581) to the strongest (GJ 832) \ion{C}{4} emitter.  This is particularly interesting in light of the established correlation between X-ray and \ion{C}{4} luminosity for M dwarf emission line (dMe) flare stars~\citep{byrne89}.   With improved atmospheric models for M dwarfs or simultaneous UV and X-ray observations of a larger sample of M dwarf planet hosts, the \ion{C}{4} flux in the HZ of low mass stars could be calibrated as a proxy for the EUV and soft X-ray emission incident on the planetary atmospheres~\citep{linsky12c}. \ion{C}{4}-based EUV irradiance estimates could provide a complementary constraint to those derived from coronal X-ray measurements~\citep{forcada11}.

Ly$\alpha$ is the brightest line in the UV spectrum of low-mass stars, although resonant scattering 
in 
the interstellar medium makes direct line profiles inaccessible, even for the nearest stars~\citep{wood00,wood05}.   As described above, Ly$\alpha$ reconstruction can be a powerful technique for determining the intrinsic stellar Ly$\alpha$ line flux, but this technique is not possible for more distant targets where the ISM completely removes stellar Ly$\alpha$ photons from our line of sight ($N$(HI)~$\gtrsim$~10$^{20}$ cm$^{-2}$).  In this case, one requires a proxy formed in a similar region of the chromosphere to estimate the local Ly$\alpha$ emission flux.  For our limited sample, we have found that Ly$\alpha$ and \ion{Mg}{2} are well correlated (see also Linsky et al. 2012a), and we present a scaling relation to infer the Ly$\alpha$ flux from a \ion{Mg}{2} observation. \nocite{linsky12c}

Figure 4 shows the reconstructed Ly$\alpha$ flux versus the observed \ion{Mg}{2} flux, scaled to 0.16 AU from the host star for GJ 581, GJ 876, GJ 436, GJ 832, and GJ 667C (black filled circles).  A linear fit to the correlation  ($F$(Ly$\alpha$) = $a$~+~$b$$F$(\ion{Mg}{2}), where $a$~=~$-$2.3~$\pm$~2.2 and  $b$~=~$+$14.8~$\pm$~1.2  ) is also displayed in Figure 4.  The observed $F$(Ly$\alpha$)/$F$(\ion{Mg}{2}) ratio is the slope of this fit.  
Chromospheric \ion{Mg}{2} emission lines of G and K dwarfs are known to show self-reversed profiles due to non-LTE effects in their upper atmospheres (e.g., Linsky \& Wood 1996; Wood \& Linsky 1998; Redfield \& Linsky 2002), however the situation is less clear for M dwarfs.  For example,~\citet{redfield02} present GHRS spectra of AU Mic, where no appreciable self-reversal of the reconstructed \ion{Mg}{2} profile is observed.  We therefore assume a simple Gaussian line shape for the M dwarf \ion{Mg}{2} emission lines.  Interstellar \ion{Mg}{2} absorption will produce the appearance of a self-reversed emission line when the stellar velocity is close to the velocity of the Mg$^{+}$-bearing interstellar cloud (e.g., Wood et al. 2000), therefore a correction for interstellar \ion{Mg}{2} absorption needs to be made in order to give an unbiased estimate of the $F$(Ly$\alpha$)/$F$(\ion{Mg}{2}) ratio for the MUSCLES stars.~\nocite{wood00,linsky96,redfield02,wood98}  

We cannot analyze the interstellar \ion{Mg}{2} absorption in detail for most of our targets because the stellar and interstellar features are unresolved in STIS G230L observations.  However, utilizing our STIS E230H high-resolution ($\Delta$$v$~$<$~3 km s$^{-1}$) spectra of GJ 832 and GJ 667C, the correction for interstellar \ion{Mg}{2} absorption can be estimated.  The profile of the \ion{Mg}{2} $k$ line ($\lambda_{o}$~=~2795.528~\AA) of GJ 832 in shown in Figure 5.  For GJ 832 and GJ 667C, we find that the    interstellar absorber removes 30~--~35\% of the intrinsic Gaussian emission line flux.  Therefore, we estimate that the scaling of the ISM-corrected \ion{Mg}{2} emission line flux to the intrinsic Ly$\alpha$ emission line flux is $\approx$~10.  Factoring in uncertainties on the interstellar correction and the Ly$\alpha$ reconstruction, we estimate that this scaling relation is good to $\sim$~30\%, or the ISM-corrected intrinsic  $F$(Ly$\alpha$)/$F$(\ion{Mg}{2}) ratio for weakly active M dwarf exoplanet host stars is 10~$\pm$~3.  Linsky et al. (2012a) find that the F(Ly$\alpha$)/F(\ion{Mg}{2}) ratios for the more active M dwarfs AU Mic and AD Leo are 2.5 and 4.0, respectively. This suggests that the ratio may decrease with activity.

\subsubsection{Non-detection of Ly$\alpha$ in GJ 1214}

GJ 1214 is the coolest (M6V, $T_{eff}$~2949~$\pm$~30 K; Kundurthy et al. 2011) and most distant source in the MUSCLES pilot study, and the only star where \ion{H}{1} Ly$\alpha$ is not detected.   In the top panel of Figure 6, we display the full UV spectrum of GJ 1214.  The only clearly detected stellar emission lines are those of \ion{C}{4} and \ion{Mg}{2}.  In the lower panel, we focus on the immediate spectral region around the lines of interest.  The upper limit from the coadded STIS G140M spectra of GJ 1214 is $F$(Ly$\alpha$)~$<$2.4~$\times$~10$^{-15}$ erg cm$^{-2}$ s$^{-1}$.
The short-wavelength component of the \ion{C}{4} doublet is shown in red, and we show a possible detection of \ion{N}{5} in blue.   Taking the observed \ion{Mg}{2} flux from GJ 1214 (2.21~($\pm$~0.15)~$\times$~10$^{-15}$ erg cm$^{-2}$ s$^{-1}$), we can use the $F$(Ly$\alpha$)/$F$(\ion{Mg}{2}) relation described above to compare the expected flux from Ly$\alpha$ with the upper limit (red circle and red star in Figure 4).   The upper limit on the GJ 1214 Ly$\alpha$ flux is an order of magnitude below the predicted point based on the observed \ion{Mg}{2} flux.  Note however that this relation is for reconstructed Ly$\alpha$ emission lines and we have no information on which to base a reconstruction in GJ 1214.   

\begin{figure}
\begin{center}
\epsfig{figure=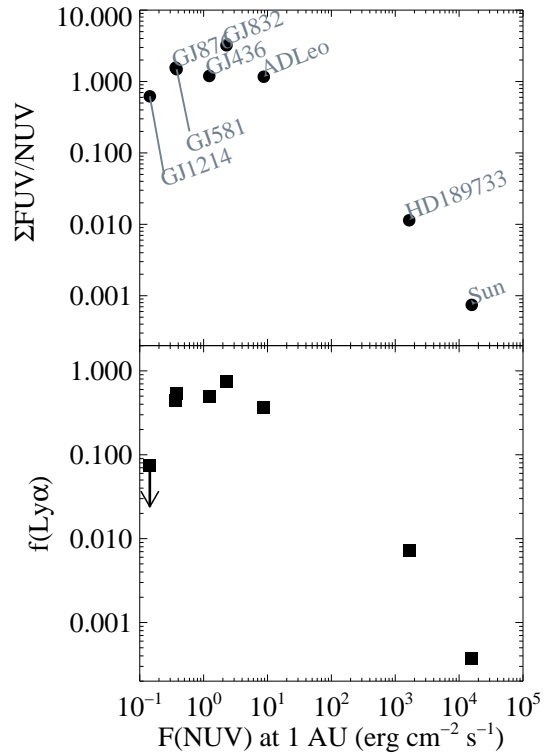,width=3.0in,angle=00}
\vspace{+0.1in}
\caption{
\label{cosovly} ($top$) The total FUV (including the reconstructed Ly$\alpha$ emission line) to NUV flux ratio is $\sim$~0.5~--~3 for M dwarfs and $\lesssim$~10$^{-2}$ for K dwarfs and earlier.  This ratio has important implications for the oxygen photochemistry on Earth-like planets orbiting low-mass stars.  ($bottom$) The fraction of the total UV energy in the Ly$\alpha$ line is $\gtrsim$~35\% for most of the M dwarfs (GJ 581, GJ 876, GJ 436, GJ 832, and AD Leo),  $<$~8~\% for GJ 1214,  and $\lesssim$~1\% for earlier-type stars.  
 }
\end{center}
\end{figure}

The presence of the UV metal line emission demonstrates that GJ 1214 maintains some basal UV flux level.  However, the non-detection of Ly$\alpha$ is surprising in light of the results presented above.  We suggest three possible explanations that might account for the non-detection.  First, the 2012 COS and STIS G230L observations may have occurred during a period of stellar activity, whereas the star may have been UV-quiet during the 2011 STIS G140M observations.  This type of ``binary'' UV activity behavior has been observed previously in the M8 flare star VB 10~\citep{linsky95}.  
Second, if the GJ 1214 Ly$\alpha$ emission line was significantly narrower than the Ly$\alpha$ lines in the other M dwarfs, then it might be possible that the somewhat longer interstellar sightline removed the signal.  However, as shown by~\citet{wood05}, neutral interstellar hydrogen columns rarely exceed log$_{10}$ $N$(HI)~=~18.5 cm$^{-2}$ for stars nearer than 50 pc.  Therefore, this second option is really that the stellar Ly$\alpha$ emission line from GJ 1214 is unusually narrow.  There is not enough information from \ion{C}{4} or \ion{Mg}{2} to make a meaningful comparison of GJ 1214 line widths to the other stars in our sample.   The third, more speculative, possibility is that for the low effective temperature of GJ 1214, a significant fraction of the available neutral hydrogen in the stellar atmosphere is locked up in molecules (mainly H$_{2}$).  As we will discuss in \S5.3, we detect H$_{2}$ emission in all of the other M dwarfs observed with COS.  We do not detect H$_{2}$ fluorescence in GJ 1214, but this could be the consequence of a feedback process driven by increasing molecular fraction.  Ly$\alpha$ is the energy source for the observed fluorescence, so as fewer atoms are available to produce Ly$\alpha$, the H$_{2}$ fluorescence process will become unobservable in FUV observations even if more molecules are present in the stellar atmosphere.

\subsection{FUV/NUV Flux Ratios and the Ly$\alpha$ Fraction}

In earlier work describing the UV spectrum of GJ 876, it was shown that the FUV/NUV flux ratio is $>$ 10$^{3}$$\times$ the solar value~\citep{france12b}.  This is a consequence of an enhanced Ly$\alpha$ line strength and a much lower NUV flux owing to the cooler effective temperature of GJ 876.  In Figure 7, we show that this is a generic property of M dwarfs.  
Here, we define the FUV flux, $F$(FUV), as the total stellar flux integrated over the 1160~--~1690~\AA\ bandpass, $excluding~Ly\alpha$. Therefore the total FUV emission is $\Sigma$FUV = ($F$(Ly$\alpha$) + $F$(FUV)), the total  FUV/NUV ratio is 

\begin{equation}
\Sigma FUV/NUV = (F(Ly\alpha) + F(FUV))/F(NUV), 
\end{equation}
and the Ly$\alpha$ fraction is defined as  
\begin{equation}
f(Ly\alpha)~=~ F(Ly\alpha) / (F(Ly\alpha)+ F(FUV) + F(NUV)), 
\end{equation} 
where $F$(Ly$\alpha$) is integrated over 1210~--~1222~\AA\ and $F$(NUV) is integrated over 2300~--~3050~\AA.  

The observed Ly$\alpha$, FUV, and NUV fluxes are presented in Table 3.  We put these values in context in Table 4 by presenting the Ly$\alpha$, FUV, and NUV luminosities as the fraction of the bolometric luminosities.  We calculate the bolometric luminosity for each star by creating a continuous spectrum from 0.115~--~2.5~$\mu$m, using the appropriate M dwarf model atmosphere~\citep{pickles98} to cover the optical and IR portions of the SED.  The model atmosphere flux is scaled to the STIS G230L data from 3050~--~3080~\AA.  The use of the Pickles models will  underestimate the bolometric luminosity somewhat (by not including the entire Rayleigh-Jeans tail into the mid-IR), however we expect that this effect is similar to or smaller than the uncertainty introduced by scaling to the NUV data, where the relative contribution of the photosphere is not clear in most cases.  

$\Sigma$FUV/NUV is in the range~$\approx$~1~--~3 for all M dwarfs, except GJ 1214.  
Mostly due to the lack of Ly$\alpha$ emission (\S5.1.1), $\Sigma$FUV/NUV is in the range 0.50~--~0.62 for GJ 1214.  While the  $\Sigma$FUV/NUV ratio is lower for GJ 1214 than for the rest of the M dwarfs in our sample, it is still significantly higher than for earlier type stars.  For HD 189733, this ratio has fallen to $\sim$~10$^{-2}$, and it is $<$~10$^{-3}$ for the Sun.  Similarly, $f$(Ly$\alpha$) spans 37~--~75~\% for most M dwarfs, is $<$~7.4\% for GJ 1214, is $\approx$ 0.7~\% for HD 189733, and is $\approx$~0.04~\% for the Sun.  AD Leo has an absolute NUV flux that is 4~--~40 times larger than for the less active M dwarfs in this survey, and the $\Sigma$FUV/NUV ratio could decrease dramatically during flares when strong Balmer continuum emission dominates the spectrum~\citep{hawley91}.  The implications of the large FUV/NUV flux ratio in M dwarfs are discussed in more detail below (\S7.1).  

\begin{deluxetable*}{lc|ccc}
\tabletypesize{\normalsize}
\tablecaption{M dwarf Broadband UV and Reconstructed Ly$\alpha$ Fluxes (ergs cm$^{-2}$ s$^{-1}$)\tablenotemark{a}. \label{lya_lines}}
\tablewidth{0pt}
\tablehead{
\colhead{Target} & \colhead{$d$ (pc)} & \colhead{$F$(Ly$\alpha$)\tablenotemark{b}}   &  \colhead{$F$(FUV)\tablenotemark{c}} & \colhead{$F$(NUV)\tablenotemark{c}}    \\
\colhead{} & &  \colhead{($\Delta$$\lambda$\tablenotemark{d}~=~1210~--~1222~\AA) } & \colhead{($\Delta$$\lambda$~=~1160~--~1690~\AA) }  &  \colhead{($\Delta$$\lambda$~=~2300~--~3050~\AA)} 
}
\startdata
GJ 581   & 6.3 & 3.0~$\times$~10$^{-13}$ & 3.6~$\times$~10$^{-14}$  &  2.3~$\times$~10$^{-13}$ \\
GJ 876   & 4.7 & 4.4~$\times$~10$^{-13}$ & 1.7~$\times$~10$^{-13}$  & 3.9~$\times$~10$^{-13}$  \\
GJ 436   & 10.2 & 3.5~$\times$~10$^{-13}$ & 3.3~$\times$~10$^{-14}$  & 2.8~$\times$~10$^{-13}$  \\
GJ 832   & 4.9 & 5.0~$\times$~10$^{-12}$  & 9.8~$\times$~10$^{-14}$  &  2.2~$\times$~10$^{-12}$  \\
GJ 667C  & 6.9  & 7.6~$\times$~10$^{-13}$ & $\cdots$  & $\cdots$  \\
GJ 1214  & 13.0 & $<$ 2.4~$\times$~10$^{-15}$  & 1.0~$\times$~10$^{-14}$  & 2.0~$\times$~10$^{-14}$  \\
\tableline\tableline
AD Leo         & 4.7  & 7.5~$\times$~10$^{-12}$  & 3.4~$\times$~10$^{-12}$  & 9.3~$\times$~10$^{-12}$  \\
HD 189733\tablenotemark{e}  & 19.5  & 7.5~$\times$~10$^{-13}$  & 4.1~$\times$~10$^{-13}$  & 1.0~$\times$~10$^{-10}$  \\
Sun           & 1 AU  & 5.9~$\times$~10$^{0}$ & 5.9~$\times$~10$^{0}$  & 1.6~$\times$~10$^{4}$ 
\enddata
\tablenotetext{a}{Flux measurements are averaged over all exposure times. } 
\tablenotetext{b}{Uncertainty on the reconstructed Ly$\alpha$ flux is estimated to be between 10~--~20\% for the stars observed with E140M (GJ 832, GJ 667C, and AD Leo) and 15~--~30\% for stars observed with G140M (GJ 581, GJ 876, GJ 436, and HD 189733).} 
\tablenotetext{c}{$F$(FUV) refers to the integrated 1160~--~1690~\AA\ flux {\it excluding Ly$\alpha$}.  Flux uncertainties for the broad band measurements are dominated by instrumental calibrations and target acquisition errors for faint sources.  The flux errors for $F$(FUV) and $F$(NUV) are estimated to be $\leq$~10\% and $\leq$~5\%, respectively. } 
\tablenotetext{d}{$\Delta$$\lambda$ is the bandpass over which the flux is integrated.  } 
\tablenotetext{e}{HD 189733 Ly$\alpha$ reconstruction made from direct STIS G140M observations. $F$(FUV) and $F$(NUV) were measured using a proxy star (HD 37394) scaled to the V-magnitude of HD 189733.  } 
\end{deluxetable*}

\begin{deluxetable*}{lc|ccc}
\tabletypesize{\normalsize}
\tablecaption{M dwarf Broadband UV Luminosity (ergs s$^{-1}$)\tablenotemark{a}. \label{lya_lines}}
\tablewidth{0pt}
\tablehead{
\colhead{Target} & \colhead{$d$ (pc)} & \colhead{ log$_{10}$ $L$(Ly$\alpha$)}   &  \colhead{ log$_{10}$ $L$(FUV)\tablenotemark{b}} & \colhead{ log$_{10}$ $L$(NUV)}    \\
\colhead{} & &  \colhead{($\Delta$$\lambda$\tablenotemark{c}~=~1210~--~1222~\AA) } & \colhead{($\Delta$$\lambda$~=~1160~--~1690~\AA) }  &  \colhead{($\Delta$$\lambda$~=~2300~--~3050~\AA)} 
}
\startdata
GJ 581   & 6.3 & 27.16 & 26.23                &  27.03 \\
GJ 876   & 4.7 & 27.07 & 26.65               &  27.01  \\
GJ 436   & 10.2 & 27.65 & 26.62                &  27.55  \\
GJ 832   & 4.9 & 28.15 & 26.44                &  27.80  \\
GJ 1214  & 13.0 & $<$25.69 & 26.31                &  26.61  \\
\tableline	
             &        &    log$_{10}$ $L$(Ly$\alpha$)/$L_{bol}$\tablenotemark{d}  &  log$_{10}$ $L$(FUV)/$L_{bol}$  &  log$_{10}$ $L$(NUV)/$L_{bol}$  \\
\tableline
GJ 581   & 6.3 & $-$4.11 & $-$5.04                &  $-$4.24 \\
GJ 876   & 4.7 & $-$4.26 & $-$4.68               &  $-$4.32  \\
GJ 436   & 10.2 & $-$4.07 & $-$5.10                &  $-$4.17  \\
GJ 832   & 4.9 & $-$3.50 & $-$5.21                &  $-$3.85  \\
GJ 1214  & 13.0 & $<$$-$6.24 & $-$5.62               &  $-$5.31 

\enddata
\tablenotetext{a}{Flux measurements are averaged over all exposure times. } 
\tablenotetext{b}{$L$(FUV) refers to the integrated 1160~--~1690~\AA\ flux {\it excluding Ly$\alpha$}.   } 
\tablenotetext{c}{$\Delta$$\lambda$ is the bandpass over which the flux is integrated.  } 
\tablenotetext{d}{$L_{bol}$ is the integral of the X-ray through IR flux of each star (\S5.2).   } 
\end{deluxetable*}

\subsection{The Ubiquity of H$_{2}$ Fluorescence: an Exoplanetary Origin?} 

We observe fluorescent H$_{2}$ emission lines in GJ 581, GJ 876, GJ 436, and GJ 832.  These features are photoexcited (``pumped'') by Ly$\alpha$ photons out of the $v$~=~2 level of the ground electronic state~\citep{shull78}. Under the assumption of thermalized rovibrational populations, this process is a signpost for 2000~--~4000 K molecular gas.  
We detect between four and eight fluorescent emission lines in each of these targets, pumped by Ly$\alpha$ through the (1~--~2) R(6) $\lambda$1215.73~\AA\  and (1~--~2) P(5) $\lambda$1216.07~\AA\ transitions.  The S/N of any of the individual lines is not sufficiently high for spectral fitting, so we coadded four to six of the brightest lines\footnote{The coadded H$_{2}$ line profiles include at least four of the following emission lines: (1~--~7)R(6) $\lambda$1500.24~\AA, (1~--~7)P(8) $\lambda$1524.65~\AA, (1~--~6)R(3) $\lambda$1431.01~\AA, (1~--~6)P(5) $\lambda$1446.12~\AA,(1~--~7)R(3) $\lambda$1489.57~\AA, and (1~--~7)P(5) $\lambda$1504.76~\AA.}.  The coadded line profiles were normalized to unity and fitted by assuming a Gaussian emission line shape that has been convolved with the 1500~\AA\ $HST$+COS line spread function~\citep{kriss11}.  The coadded fluorescence lines and fits are shown in Figure 8.   The H$_{2}$ emission lines are consistent with the stellar radial velocities to within half of the 15 km s$^{-1}$ wavelength solution uncertainty of the COS G160M mode, [$v_{H2}$(581)~=~$-$5.5 km s$^{-1}$, $v_{H2}$(876)~=~$+$9.0 km s$^{-1}$, $v_{H2}$(436)~=~$+$17.5 km s$^{-1}$, $v_{H2}$(832)~=~$+$19.7 km s$^{-1}$] compared with [$v_{rad}$(581)~=~$-$9.5 km s$^{-1}$, $v_{rad}$(876)~=~$+$8.7 km s$^{-1}$, $v_{rad}$(436)~=~$+$10 km s$^{-1}$, $v_{rad}$(832)~=~$+$18.0 km s$^{-1}$].  Except in the case of GJ 581 (the lowest S/N H$_{2}$ spectrum), the H$_{2}$ emission features are spectrally unresolved.   In this subsection, we describe three possible physical origins for the observed fluorescence: 1) the stellar photospheres, 2) the atmospheres of the planets orbiting these stars, or 3) circumstellar gas disks populated by atmospheric mass-loss from the planets in these systems.  

\begin{figure*}
\begin{center}
\epsfig{figure=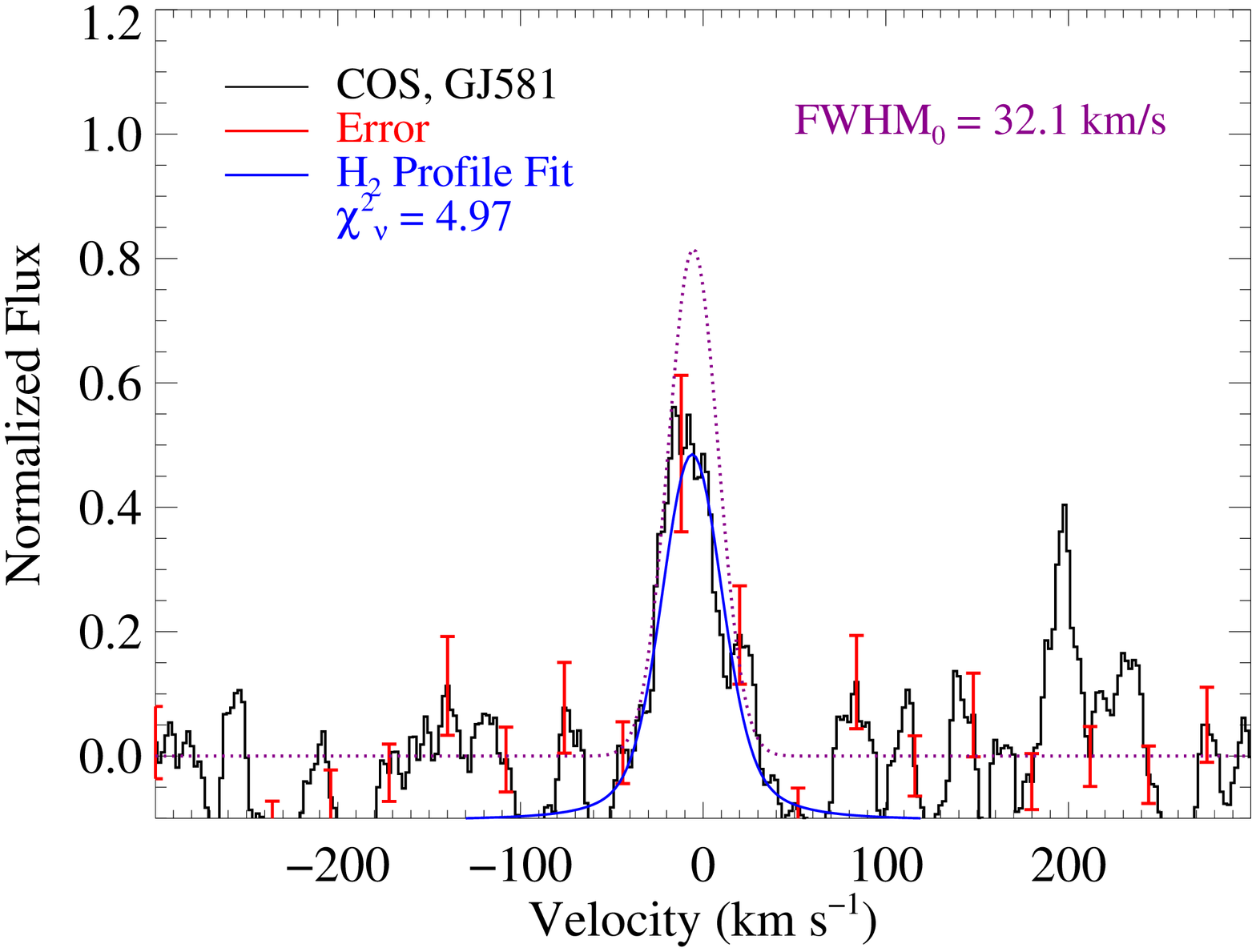,width=2.25in,angle=90}
\epsfig{figure=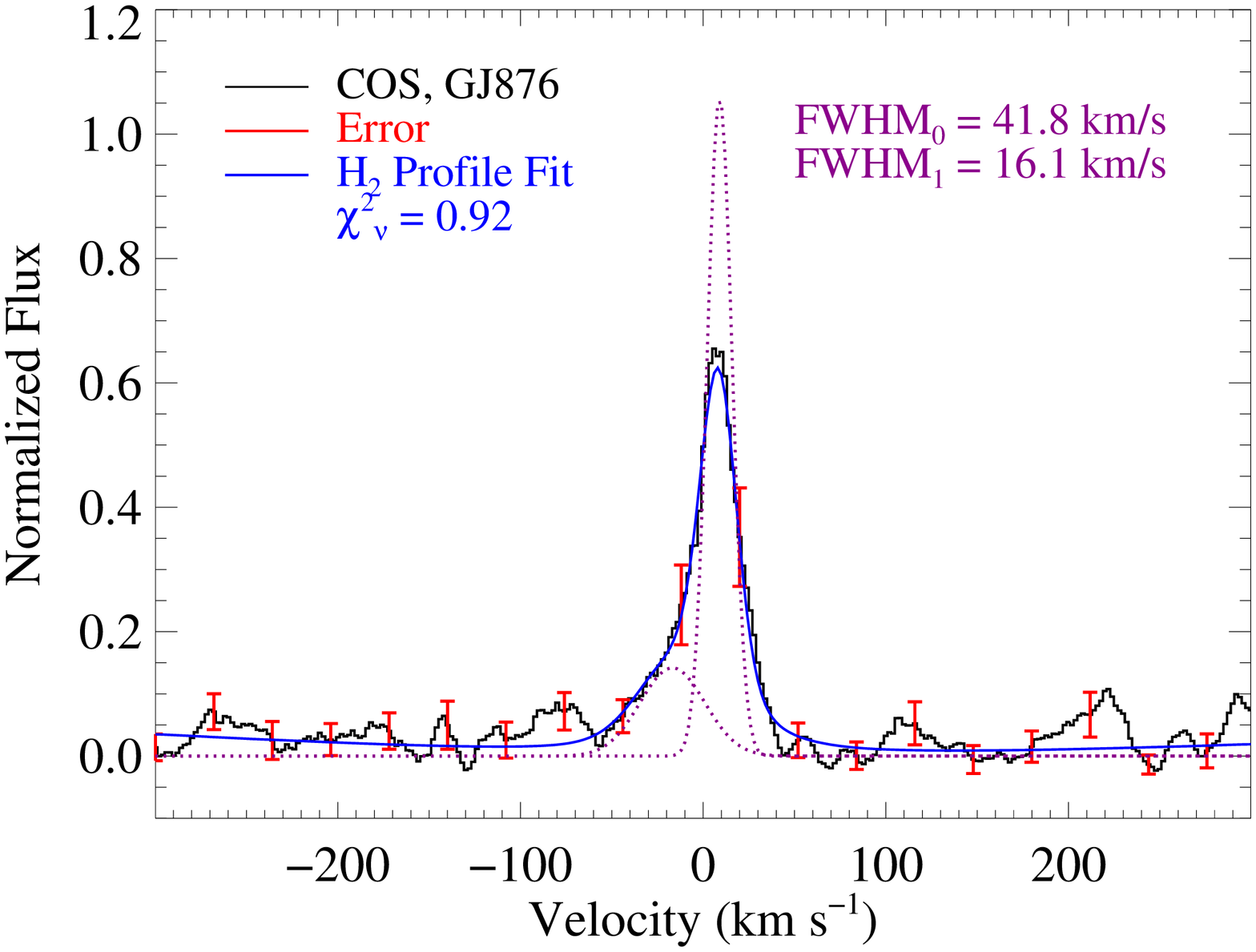,width=2.25in,angle=90}
\epsfig{figure=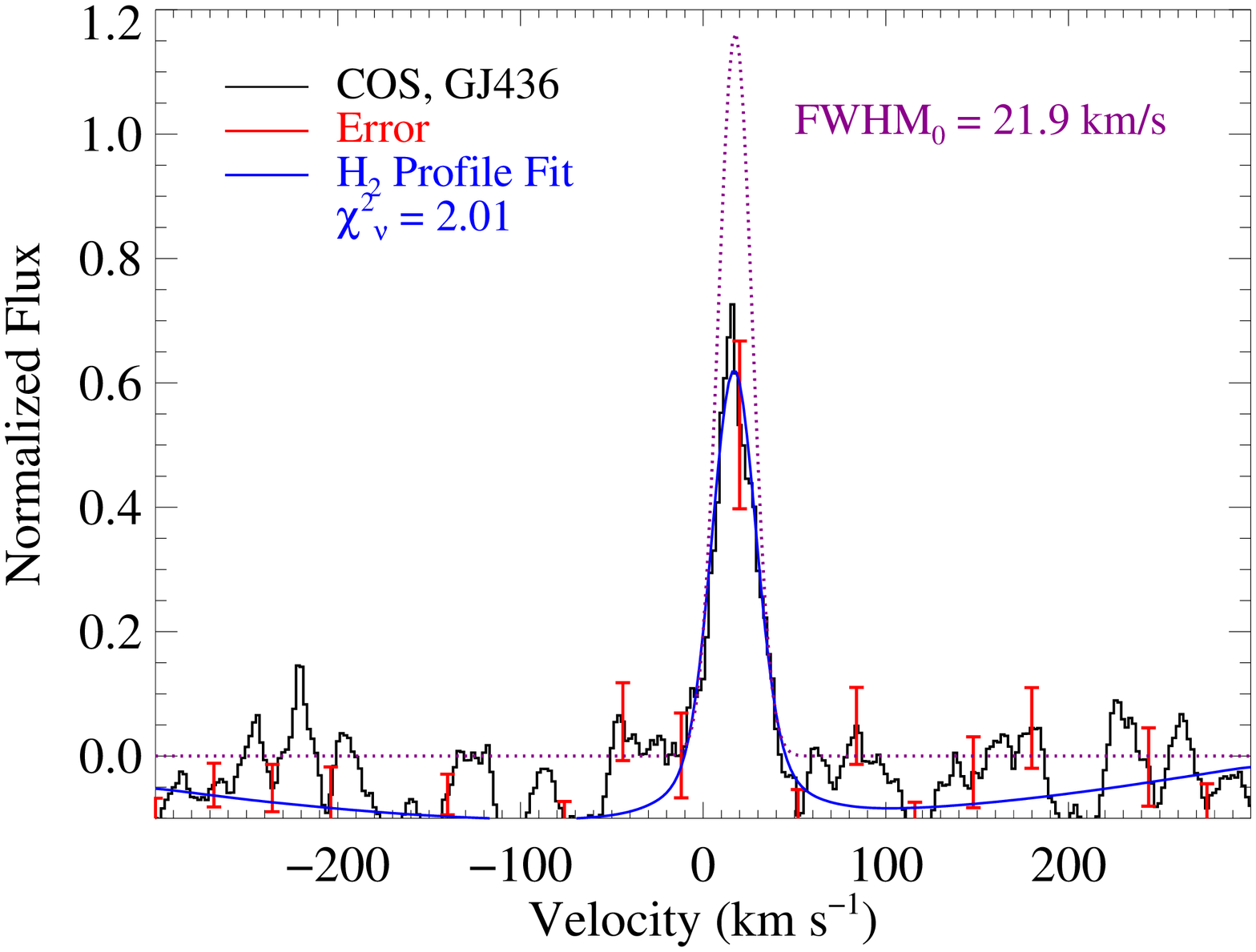,width=2.25in,angle=90}
\epsfig{figure=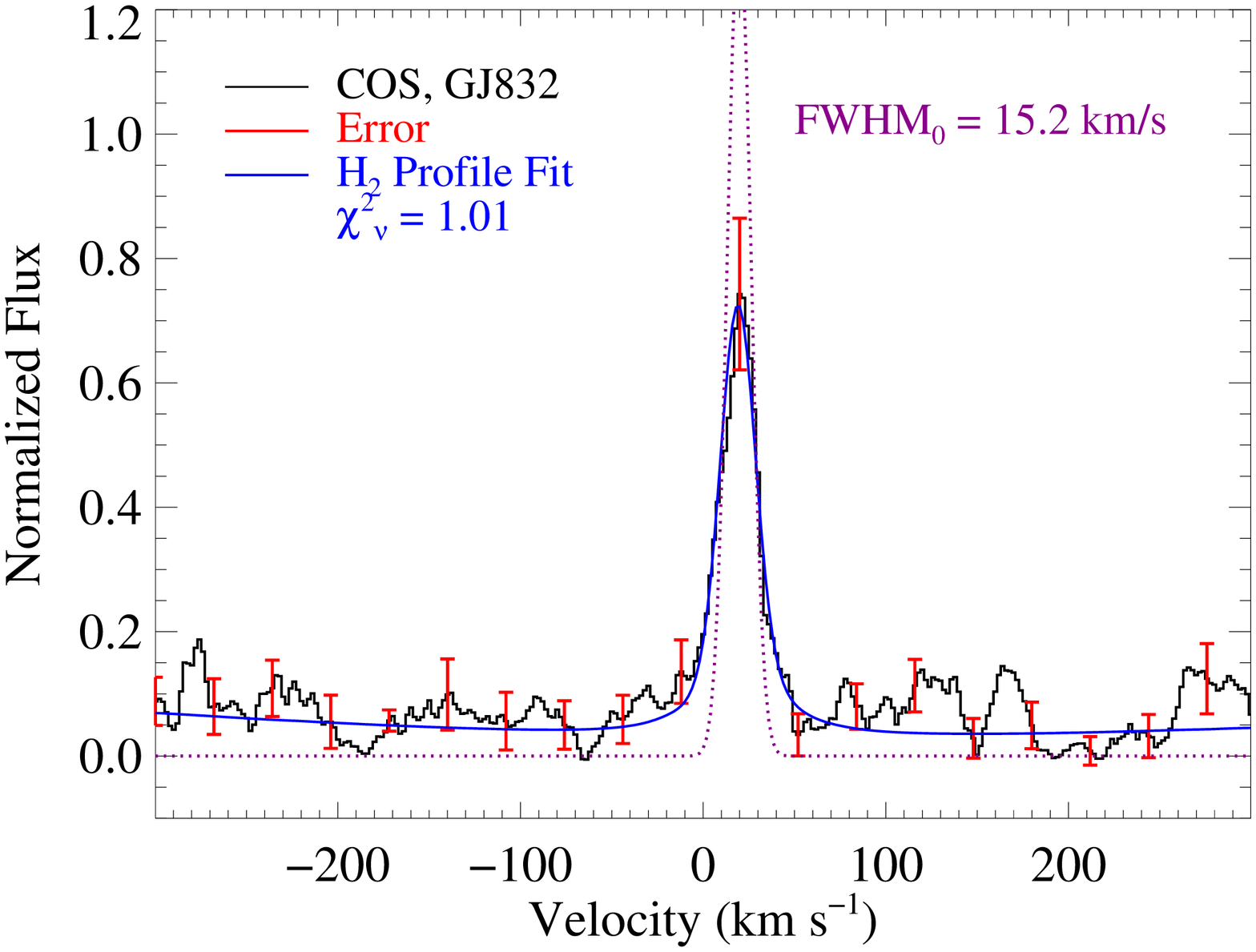,width=2.25in,angle=90}
\vspace{+0.0in}
\caption{
\label{cosovly} Coadded H$_{2}$ emission spectra detected in the MUSCLES pilot program.  Each velocity profile is the coaddition of four to six fluorescent emission lines.  The dotted line is the intrinsic Gaussian line profile which has been convolved with the COS line spread function ({\it solid blue line}) for comparison with the data ({\it black histogram}).  The red error bars are representative of the uncertainty on the normalized line profiles.  The origin of these lines is not yet clear; more detailed modeling is required to differentiate between a photospheric and exoplanetary origin for this hot, photoexcited molecular gas. 
 }
\end{center}
\end{figure*}

Ly$\alpha$-pumped H$_{2}$ emission lines were first detected in sounding rocket spectra of sunspots~\citep{jordan77}, and have been modeled assuming a thermalized H$_{2}$ population at a temperature of 3200 K~\citep{shull78}.  H$_{2}$ fluorescence lines are also powerful diagnostics of the molecular surface layers of protoplanetary disks over a similar range of temperatures~\citep{herczeg04,france12c}.  These lines are not clearly detected in previous STIS observations of M dwarfs $without$ planets (AD Leo, Proxima Cen, EV Lac, and AU Mic\footnote{AU Mic displays H$_{2}$ fluorescence from lower temperature molecular gas, but this gas is consistent with an origin in a remnant protoplanetary gas disk~\citep{france07}.}), although this could be related to much larger instrumental backgrounds associated with the STIS echelle modes.    In any event, the 3200 K sunspot temperature is characteristic of the photospheric effective temperature of the MUSCLES stars, and simple energy budget calculations (e.g., France et al. 2012a) suggest that the 
fluorescence may be consistent with an origin in a hot molecular stellar surface layer.  

However, it is interesting that these lines are either not detected or are marginally detected with much lower H$_{2}$/atomic line ratios in the non-planet hosting stars.  H$_{2}$ is the primary constituent of gas giant planets. After \ion{H}{1} Ly$\alpha$,  the electronically excited emission spectrum of H$_{2}$  dominates the FUV spectrum of giant planets~\citep{broadfoot79,wolven98,gustin10}.   Fluorescence driven by solar Ly$\beta$ is the main constituent of the dayglow (non-auroral illuminated disk) spectrum of Jupiter~\citep{feldman93} and in areas where the Jovian surface temperature is elevated (e.g., the Shoemaker-Levy 9 impact site; Wolven et al. 1997), Ly$\alpha$-pumped fluorescence becomes strong.

All four of the stars showing strong H$_{2}$ fluorescence host planets of Neptune~to~super-Jupiter mass.  Our analysis cannot rule out the possibility that the ubiquitous Ly$\alpha$-pumped H$_{2}$ fluorescence observed towards M dwarf planet hosts is the result of photoexcited dayglow emission in planetary atmospheres.  In light of the recent tentative detections of FUV H$_{2}$ emission from the planetary systems around HD 209458~\citep{france10a} and brown dwarf 2MASS J12073346-3932539~\citep{france10b}, it is tempting to attribute an exoplanetary origin to these lines.  We caution that 
 further study is required to differentiate between these two scenarios.  If the H$_{2}$ in the GJ 436 system originates in the atmosphere of GJ 436b, we would have expected to see a significant velocity shift of the fluorescent lines relative to the stellar radial velocity ($v_{436b}$~=~110~--~120 km s$^{-1}$ at orbital phase $\approx$~0.8).  However, the H$_{2}$ lines are consistent with the radial velocity of the star, indicating little direct dayglow and/or auroral contribution from GJ 436b.  The upper limit on the H$_{2}$ (1~--~7)R(3) emission line flux at +110 km s$^{-1}$ (assuming unresolved emission lines) is $\approx$~6~$\times$~10$^{-17}$ erg cm$^{-2}$ s$^{-1}$ in the GJ 436 spectrum.  
Modeling of both the H$_{2}$ fluorescence spectrum from M dwarf photospheres and the expected UV emission spectrum from nearby M dwarf planetary systems would be extremely valuable.  

A final possibility is that the observed H$_{2}$ fluorescence originates in a circumstellar gas envelope that is being replenished by atmospheric mass-loss from escaping planetary atmospheres.  The narrow H$_{2}$ line-width in GJ 436 rules out the possibility that the fluorescing gas is confined to the semi-major axis of the planetary orbit ($a$~=~0.03 AU for GJ 436b), however over the $\sim$~5 Gyr lifetimes of these systems, this material could disperse into a low density ``second generation'' circumstellar disk.  While detailed modeling of such a disk is beyond the scope of this work, we can make a rough estimate of the planetary mass-loss rate required to sustain the observed H$_{2}$ emission.  

The balance equation of H$_{2}$ photodissociation and replenishment is $\beta_{diss}$$n$(H$_{2}$)~=~$\Phi_{loss}$.
Due to the negligible 912~--~1120~\AA\ continuum flux from inactive M-type stars,   
we assume that the interstellar radiation field (ISRF) dominates H$_{2}$ dissociation.  The photodissociation rate is $\beta_{diss}$~=~$B_{diss}$$\beta_{o}$.  $B_{diss}$ is the dissociation fraction of molecules excited into the Lyman and Werner bands following absorption of a 912~--~1300~\AA\ photon, $\approx$~0.2 for a hot H$_{2}$ gas.  $\beta_{o}$ is the standard H$_{2}$ photoabsorption rate from the ISRF, $\beta_{o}$~=~5~$\times$~10$^{-10}$ s$^{-1}$~\citep{jura75a}.  We assume an H$_{2}$ column density of $N$(H$_{2}$)~$\approx$~10$^{17}$ cm$^{-2}$~\citep{france12b} and a fiducial cloud diameter of $d_{H2}$ = 1 AU such that $n$(H$_{2}$)~=~$N$(H$_{2}$)/$d_{H2}$~$\approx$~7~$\times$~10$^{3}$ cm$^{-3}$.  
The replenishment term from the escaping atmospheres of our short-period planets is   $\Phi_{loss}$~=~{\it \.{M$_{loss}$}}$~m_{H2}^{-1}$~$F$(H$_{2}$/HI)~$V_{emit}^{-1}$, where {\it \.{M$_{loss}$}} is the planetary atmosphere mass-loss rate, $F$(H$_{2}$/HI) is the fraction of the total cloud in molecular form, and $V_{emit}$ is the volume of the emitting area ($V_{emit}$~=~$\pi$$r_{H2}^{2}$$z_{H2}$).  Models of escaping hot Jupiter atmospheres find that most of the escaping material is atomic~(e.g., Tremblin \& Chiang 2012), so we assume a low molecular fraction $F$(H$_{2}$/HI)~=~10$^{-3}$.\nocite{tremblin12}  Any disk populated by an orbiting planet will most likely be geometrically thin, and we take $z_{H2}$~=~0.01 $d_{H2}$ as the vertical height.  Keeping the crudeness of this approach in mind, we find that the H$_{2}$ disk can be sustained for mass-loss rates {\it \.{M$_{loss}$}}~$\gtrsim$~6~$\times$~10$^{10}$ g s$^{-1}$, well within the range of mass-loss estimates of short-period planets around G, K, and M-type stars~\citep{vidal08,linsky10,lecavelier12,ehrenreich11,ehrenreich11b}.  Therefore, the H$_{2}$ fluorescence observed in our MUSCLES sample appears compatible with either a photospheric or circumstellar origin.  

\begin{figure*}
\begin{center}
\epsfig{figure=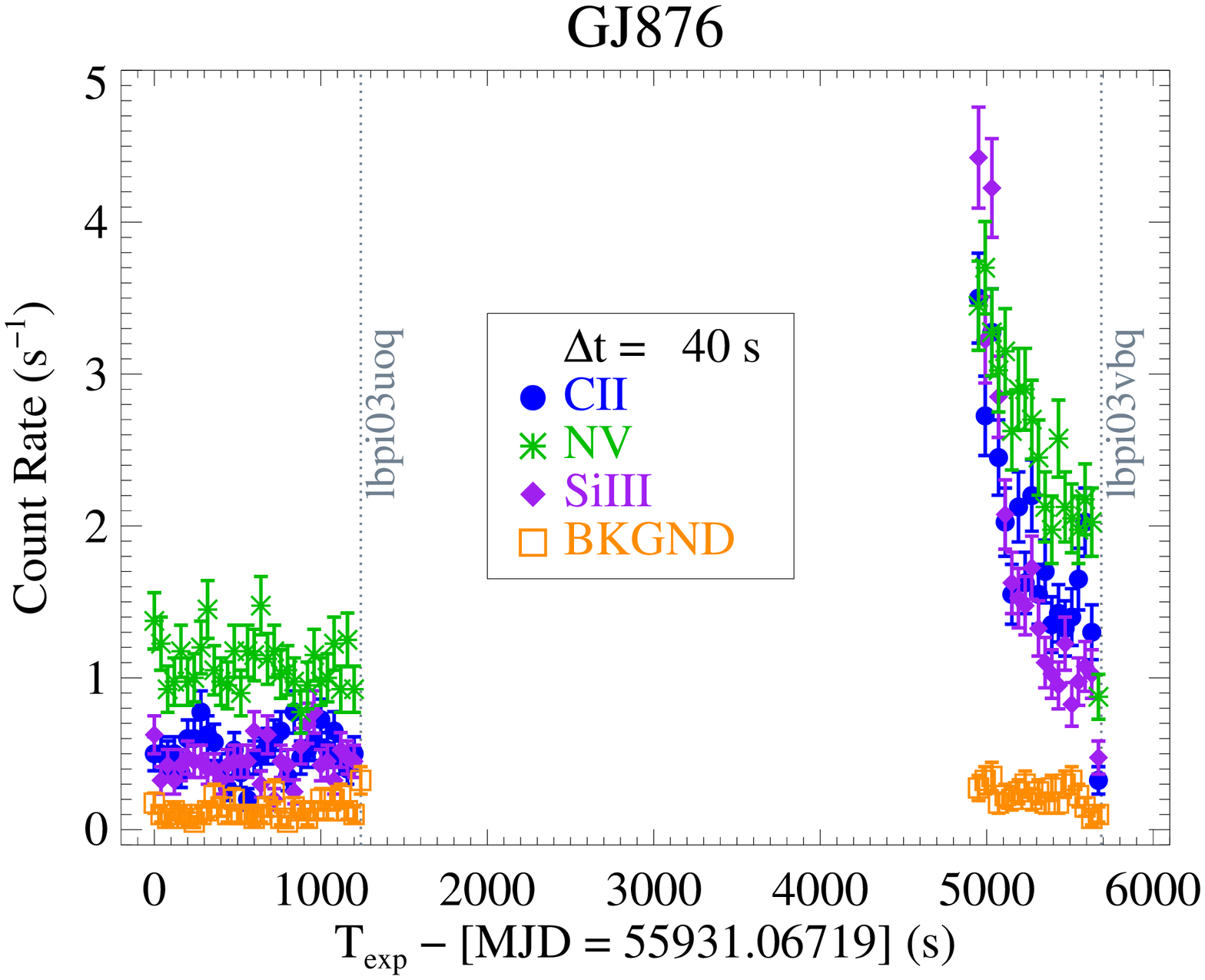,width=2.25in,angle=90}
\epsfig{figure=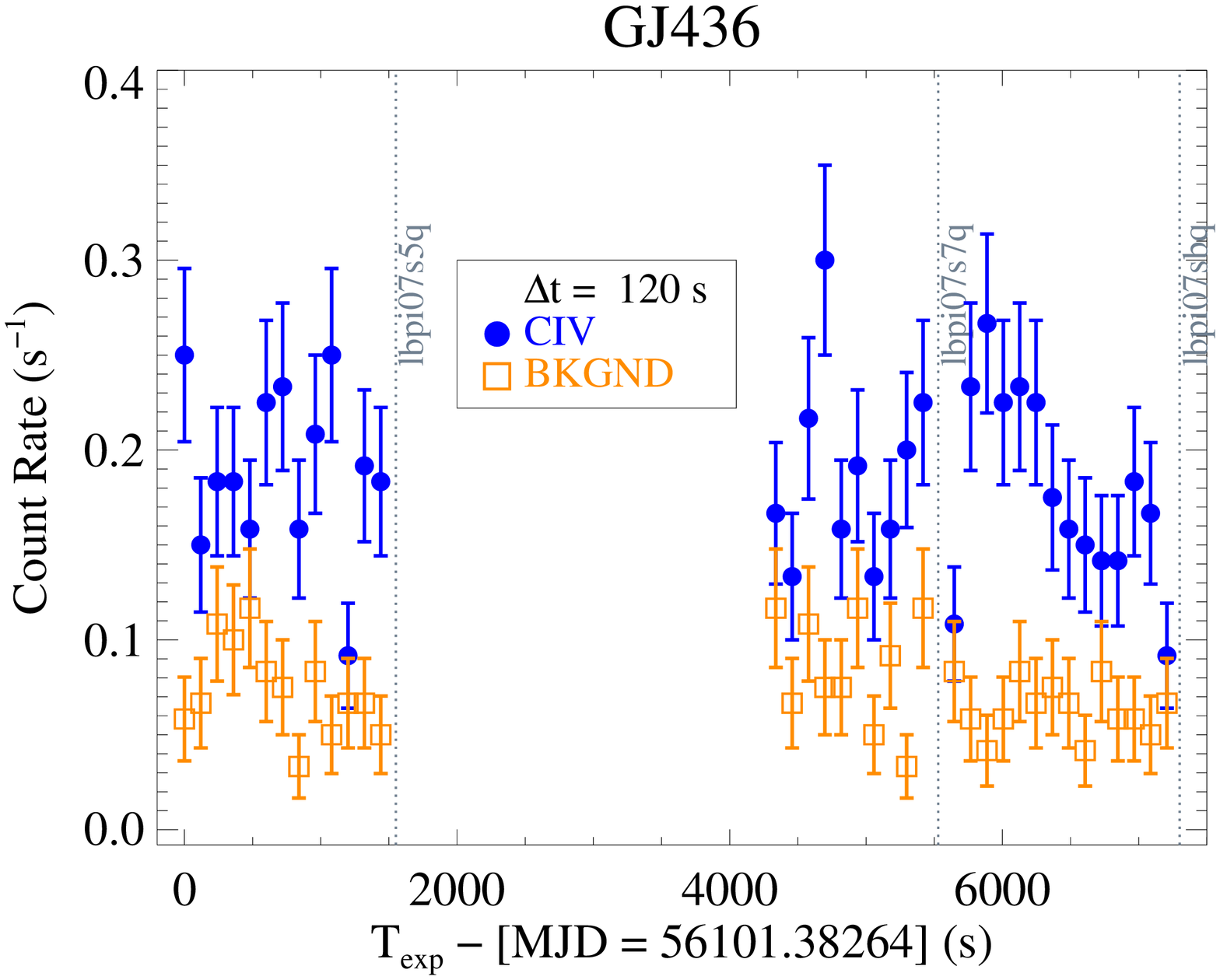,width=2.25in,angle=90}
\epsfig{figure=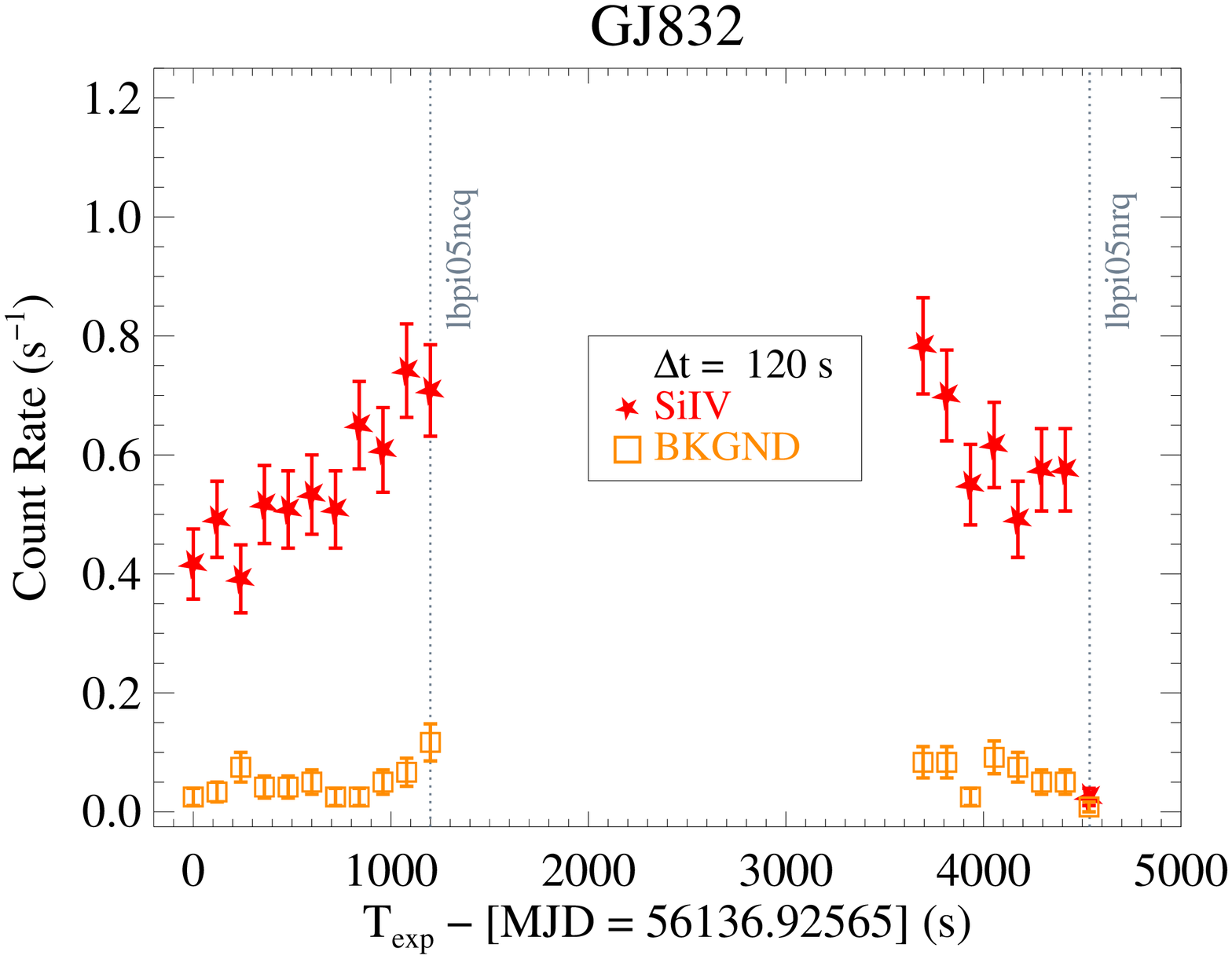,width=2.25in,angle=90}
\epsfig{figure=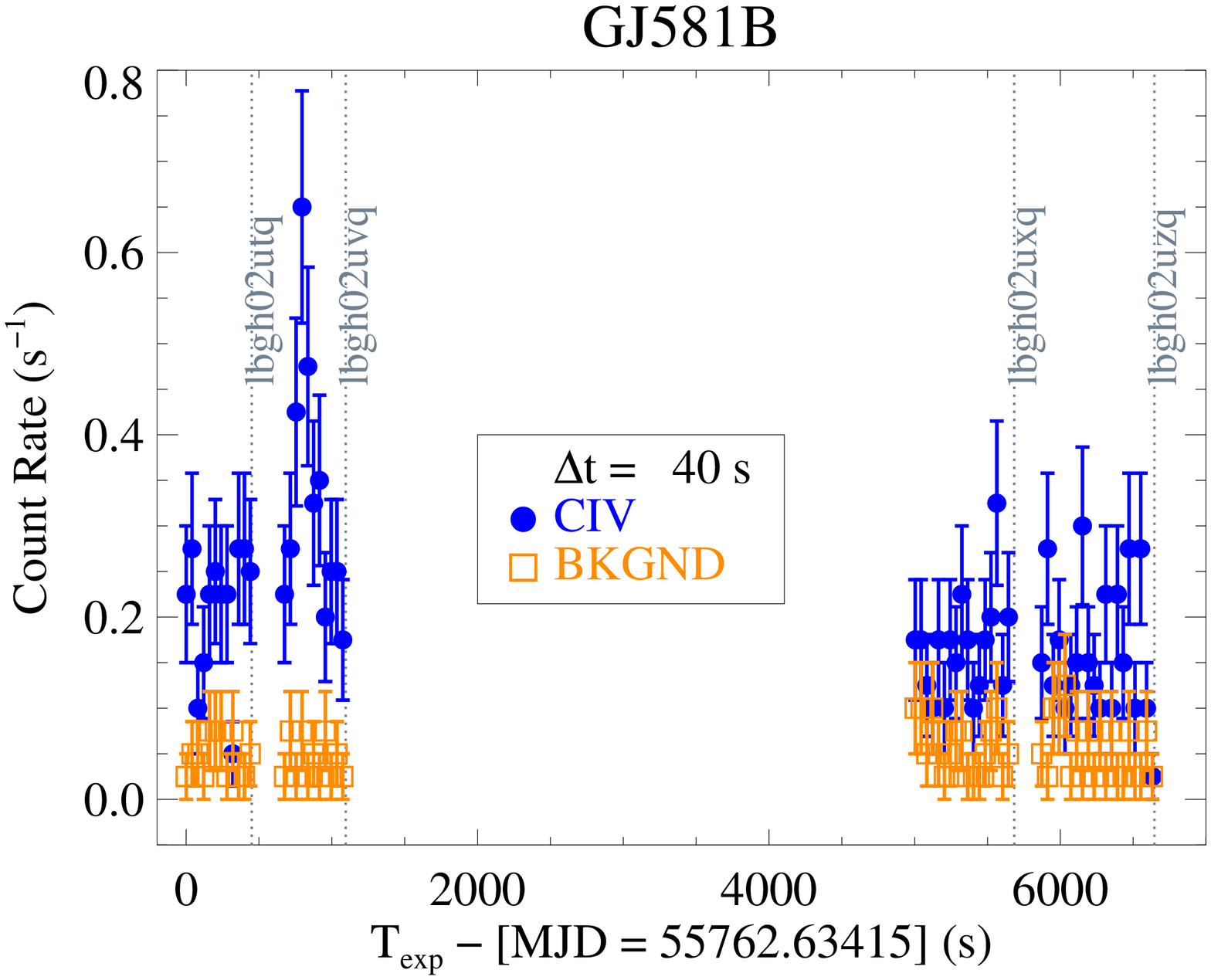,width=2.25in,angle=90}
\vspace{+0.0in}
\caption{
\label{cosovly} We have detected a range of temporal variability scales in the chromospheric and transition region emission lines from weakly active M dwarf exoplanet host stars.  We separate the time variability into three qualitative groups:  $(i)$ The Impulsive Flare, shown in several chromospheric and transition region tracers (\ion{C}{2}, \ion{Si}{3}, \ion{C}{4}, and \ion{N}{5}) in GJ 876 and GJ 436 (top row).  $(ii)$ The Slow Rise Flare, shown in \ion{Si}{4} in GJ 832 (lower left).  $(iii)$ The Short Flare, shown in \ion{C}{4} in GJ 581.  This demonstrates that time-variable, active atmospheres are a generic property of M dwarf exoplanet host stars. The gray vertical dotted lines denote the end of an individual $HST$-COS exposure.  Exposure IDs are included.  
 }
\end{center}
\end{figure*}

\section{Time Variability of M dwarf UV Radiation Fields}

We show examples of the UV light curves of GJ 581, GJ 876, GJ 436, and GJ 832 in Figure 9.  The gaps in the light curves shown in Figure 9 are due to Earth occultations during the orbit of $HST$.  We observe time-variability in the UV spectra of all of our M dwarf exoplanet host stars observed with COS (GJ 1214 has insufficient S/N for timing analysis), even on weakly active stars~\citep{walkowicz09}.  In Figure 9, we show example light curves from all four targets in chromospheric and transition region tracers \ion{Si}{3} $\lambda$1206, \ion{N}{5} $\lambda$$\lambda$1239,1243, \ion{C}{2} $\lambda$$\lambda$1334,1335, \ion{Si}{4} $\lambda$$\lambda$1394,1403, and \ion{C}{4} $\lambda$$\lambda$1548,1550, and the detector background, evaluated in $\Delta$$t$ = 40s or 120s timesteps.  Our observations were designed to provide a first estimate of the flux levels and temporal variability in low-mass exoplanet host stars, so the observations do not span a sufficient baseline to determine the UV flare frequency. However, we can qualitatively identify three flare behaviors in our spectra:

{\it 1) The Impulsive Flare}~--~The GJ 876 and GJ 436 UV flares exhibit a qualitatively similar  behavior to optical light flares on M dwarfs (e.g., Figure 5 of Kowalski et al. 2009).\nocite{kowalski09}  We observed a flare peak/quiescent flux ratio of $\geq$~10 on GJ 876, with an approximately exponential decay ($\tau_{876}$~$\approx$~400s).   The UV flare behavior on GJ 876 is similar to the factor of $\sim$~5~--~20 flux increase observed on AD Leo~\citep{hawley03}.   GJ 876 also displays X-ray flare activity~\citep{poppenhaeger10}.   A similar but lower amplitude \ion{C}{4} flare was observed on GJ 436.  The \ion{C}{4} flux approximately doubled in 240s, then decayed to pre-flare levels on a timescale $\tau_{436}$~$\approx$~1200s.

{\it 2) The Slow Rise Flare}~--~The \ion{Si}{4} light curve of GJ 832, binned to $\Delta$$t$~=~120s timesteps, is shown in the lower left hand panel of Figure 9.  Compared with the abrupt flux increase seen in GJ 876 and GJ 436, the GJ 832 \ion{Si}{4} flux increases by a factor of 1.5~--~2.0 over the course of the first 1200s exposure.  The peak and turnover are lost to Earth occultation between COS exposures for GJ 832, but the flare is observed to slowly decay during the last 960s of G130M observation.  

{\it 3) The Short Flare}~--~GJ 581 displays a shorter duration flare event, shown in the lower right panel in Figure 9.  The \ion{C}{4} light curve shows a flux increase of a factor of $\sim$~2~--~3 at $T_{exp}$~$\approx$~750s.  The signal-to-noise in the GJ 581 UV emission lines is low, but one can resolve the flare using $\Delta$$t$~=~40s timesteps.  This 200s long \ion{C}{4} emission enhancement comes during the first $HST$ orbit, when the basal transition region flux was approximately a factor of two larger than during the second $HST$ orbit following Earth occultation.  

The conclusion from this section is that even with a small sample of relatively low S/N spectra obtained over limited temporal baselines, we have identified variability with a range of amplitudes and durations. Continuous $HST$ observing campaigns, spanning 6~--~10 orbits each, of a larger sample of exoplanet host stars will be essential to better constraining the frequency, duration, and amplitude of the variable UV radiation environment of planets around M dwarfs. 
Alternatively, long-duration NUV observations could be made with balloon-borne experiments.  Time-domain studies at optical wavelengths are also of interest for characterizing the flare-frequency of the largest flares that are panchromatic events~(e.g., Hawley et al. 2003).   This is particularly relevant in light of the projected paucity of UV-capable observatories.  A more complete understanding of UV variability on weakly active and inactive M dwarfs, in combination with better constrained flare frequencies for stars of a variety of activity strengths, would allow one to use the wealth of data from current and upcoming large optical surveys to inform planetary atmosphere models.

\section{Discussion}

\subsection{UV Radiation Fields for Atmospheric Photochemistry Models}

Owing to the scarcity of M dwarf exoplanet host spectra previously available in the literature, authors modeling the atmospheres of Earth-like planets around M dwarfs have typically been driven to one of two extremes: assuming no UV flux~\citep{segura05,rauer11,kaltenegger11} or assuming the spectrum of one of the most active M dwarfs known, AD Leo~\citep{segura05,wordsworth10}.  In the case of no UV flux, the short wavelength spectral cut off is determined by the photosphere of a theoretical star with $T_{eff}$~$\lesssim$~3500 K, typically having negligible flux below 2500~\AA\ and no chromospheric emission features.  \citet{walkowicz08} demonstrated that even M dwarfs with modest X-ray emission and no chromospheric H$\alpha$ emission display NUV spectra that are qualitatively similar to more optically active stars.  Following this work, \citet{kaltenegger11} noted that an active UV atmosphere is most likely present on almost all M dwarfs, which is consistent with the observations presented here.  Additional observations are required to understand the UV flux from M dwarfs as a function of stellar age and metallicity, for spectral types later than M6, and for a larger sample of M dwarfs without X-ray emission or H$\alpha$ activity.  

The MUSCLES observations show that the UV spectral behavior of four of our stars (GJ 581, GJ 876, GJ 436, and GJ 832) is similar to AD Leo in both the spectral and the temporal sense.  The MUSCLES stars (except GJ 1214) showed similar a $\Sigma$FUV/NUV flux ratio and $f$(Ly$\alpha$) to that of AD Leo, and they also displayed time variability with UV emission line amplitudes changing by factors of $\gtrsim$~2 during our brief observing snapshots.  

While AD Leo appears to be a more appropriate choice for exoplanetary modeling input, a primary difference is the absolute flux of the NUV radiation field from AD Leo.  AD Leo has factors of $\sim$~4~--~40 stronger absolute flux levels at its effective HZ distance than do the rest of the MUSCLES targets.  Furthermore, the strong Balmer continuum associated with its frequent flare activity will periodically drive the $\Sigma$FUV/NUV ratio significantly below the level ($\sim$~1~--~3) found for the less active M dwarfs.  This could lead to significant differences between the atmospheric  chemistry of a planet orbiting an AD Leo-like star and one orbiting a more typical M dwarf.  
It has been recently shown that the strong FUV flux and high $\Sigma$FUV/NUV flux ratio incident on a terrestrial planet orbiting in the HZ of an M dwarf can lead to the buildup of a large, abiotic atmospheric O$_{2}$ abundance through the photodissociation of CO$_{2}$~\citep{tian12}.  The creation of O$_{3}$ via O$_{2}$ photolysis depends sensitively on the strength of the 1700~--~2400~\AA\ radiation field and the balance of the ocean-atmosphere system on these planets.  In some models, this process can lead to detectable levels of O$_{3}$ without the presence of an active biosphere~\citep{goldman12}.   

Spectral observations of O$_{2}$ and O$_{3}$, along with CH$_{4}$ and CO$_{2}$, are expected to be the most important signatures of biological activity on planets with Earth-like atmospheres~\citep{marais02,seager09}, which may  be the case for such planets orbiting solar-type stars.  However, the above work has shown that when realistic inputs for the UV radiation field are included, it may be possible to produce O$_{2}$ and O$_{3}$ with observable abundances on terrestrial planets in the HZ of M dwarfs {\it without the presence of biological life}.  Therefore, a detailed knowledge of the stellar spectrum will be critical to interpreting the observations of potential biomarkers on these worlds when they are detected in the coming decades.  Until such time that stellar models of M-type stars can reliably predict the observed UV-thru-IR spectrum, a direct UV observation will be the most reliable means for determining the radiation field incident on planets orbiting M dwarfs.   

\begin{figure}
\begin{center}
\epsfig{figure=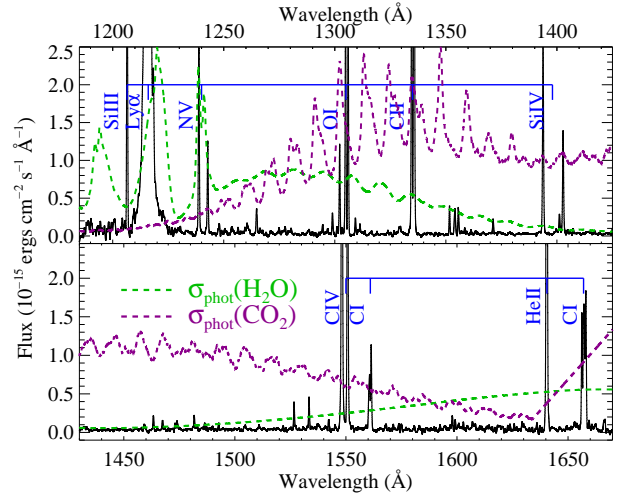,width=2.6in,angle=90}
\vspace{-0.1in}
\caption{
\label{cosovly} Photo-absorption cross-sections of H$_{2}$O (green dashed line; Mota et al. 2005) and CO$_{2}$ (purple dashed line; Yoshino et al. 1996) showing the spectral overlap with chromospheric and transition region emission lines from an M dwarf (black histogram).  Bright chromospheric and transition region emission lines are labeled.
}
\end{center}
\end{figure}

\subsection{UV Transit Spectroscopy of M dwarf Planets}

In Section 6, we described the range of time-variability observed in the UV chromospheric and transition region 
emission lines from our M dwarf exoplanet host stars.  We observe emission line flux variability of 50~--~500\% on timescales shorter than the typical transit duration of short-period exoplanets (few hours).  
For comparison, the typical depth of a resolved UV transit observation is $\sim$~5~--~15\%~\citep{vidal03,jaffel07,linsky10,lecavelier12}.  
The importance of stellar variability for UV transit studies of G and K dwarfs is just now being explored~\citep{haswell12}.  While an older solar-type transiting planet host star such as HD 209458 has been shown to have very steady transition region emission (observed in \ion{Si}{4}; Linsky et al. 2010), UV variability on younger, less-massive transiting planet hosts like HD 189733 may significantly hinder the ability to detect transits and characterize their depth and ingress/egress profiles.  GJ 436 and GJ 1214 both host transiting planets (Neptune and super-Earth mass planets, respectively), and our results suggest that transit studies using UV chromospheric emission lines as background sources to search for signs of atmospheric escape may be complicated by stellar activity.  NUV searches for O$_{3}$ in the atmospheres of Earth-like planets will also likely be impacted by M dwarf stellar activity, even for weakly active stars.  Long-baseline observations over multiple transits will be required to mitigate the effects of stellar activity in future M dwarf exoplanet transit observations.  We will consider this issue in more detail in a future work.  

\subsection{The MUSCLES Spectral Database} 

As part of the MUSCLES pilot program, we have started building a library of medium/high resolution UV spectra of M dwarf exoplanet host stars. These data are available to download for use in photochemical models of  Earth-like, ice giant, and gas giant planets\footnote{The combined 1150~--~3100~\AA\ spectra (0.05~\AA\ pixel$^{-1}$ binning) and the full-resolution, coadded COS spectra of the MUSCLES targets are available for download at: {\tt http://cos.colorado.edu/$\sim$kevinf/muscles.html}.  }.  

The high-resolution of the MUSCLES sample is important for detailed molecular emission/absorption modeling.  As most of the UV flux from low-mass stars arises in discrete emission lines, molecular species in the atmospheres of HZ planets will be ``selectively pumped'': only species that have spectral coincidences with stellar emission lines will be subject to large energy input from the host star.   H$_{2}$O, CH$_{4}$, and CO$_{2}$ in particular are highly sensitive to far-UV radiation, and provide an illustration of the utility of spectrally resolved data. CH$_{4}$ absorbs in a broad band at $\lambda$~$<$~1500~\AA, and the photoabsorption rates are relatively insensitive to the spectral resolution.  CO$_{2}$ displays a banded structure over a broad absorption region from $\sim$~1200~--~1600~\AA, and high resolution data show a spectral coincidence between the narrow CO$_{2}$ bands and the \ion{Si}{3} $\lambda$1298  + \ion{O}{1} $\lambda$1304 multiplet and the \ion{C}{2} $\lambda$1335 stellar emission lines.  H$_{2}$O dissociation is influenced by far-UV photons as the molecules absorb in a broad continuum from 1400~--~1900~\AA, $B$~--~$X$ absorption bands near 1280~\AA, and sharp Rydberg transitions at $\lambda$ $<$ 1240~\AA\ that overlap with the Ly$\alpha$ emission line~\citep{yi07}.   The spectral overlap between the photo-absorption cross-sections of CO$_{2}$ and H$_{2}$O and the chromospheric and transition region emission lines from an M dwarf are shown in Figure 10.  Low-resolution spectra or molecular absorption cross-sections will fail to preserve the specific energy absorption in these narrow ($\Delta$$\lambda$~$<$~10~\AA) wavelength coincidences and could lead to an underestimation of the dissociation rates of important atmospheric species.

\section{Summary}

We have presented the first study of the UV radiation fields around nearby M dwarf planet hosts that covers both FUV and NUV wavelengths.  Using a combination of data from the COS and STIS instruments on $HST$, we present 1150~--~3100~\AA\ spectra for GJ 581, GJ 876, GJ 436, GJ 832, GJ 1214, and Ly$\alpha$ and \ion{Mg}{2} observations for GJ 667C.  Using an iterative least-squares approach, we reconstruct the intrinsic Ly$\alpha$ radiation field strengths assuming a two-component Gaussian emission line.  Using the time-tagged capability of the COS FUV detector, we calculated emission line light curves in several chromospheric and transition region diagnostics.  The main results of this work are: 
\begin{enumerate}

	\item All six of the stars in our sample display both FUV and NUV emission line flux, even though these stars do not display H$\alpha$ emission (however, they do show some degree of H$\alpha$ absorption and \ion{Ca}{2} emission lines).  This suggests that an active stellar UV atmosphere should be included in calculations of atmospheric chemistry on planets orbiting most M dwarfs.

	\item The Ly$\alpha$ emission lines are well-fit by a two-component Gaussian emission line model with narrow (FWHM$_{narrow}$~$\sim$~90~--~130 km s$^{-1}$) and broad (FWHM$_{broad}$~$\sim$~160~--~300 km s$^{-1}$) components.  The $F_{narrow}$/$F_{broad}$  flux ratio ranges from $\sim$~5~--~12.  Interstellar \ion{H}{1} column densities are derived; 18.06~$\leq$~log$_{10}$$N$(HI)~$\leq$~18.47 cm$^{-2}$ for our targets ($d$~=~4.7~--~10.3 pc). 

	\item Ly$\alpha$ emission was not detected towards the M6 super-Earth host star GJ 1214, although emission from \ion{Mg}{2} and \ion{C}{4} indicate that an active atmosphere is present, at least during certain times.   

	\item The reconstructed Ly$\alpha$ line flux correlates with the observed \ion{Mg}{2} line flux.  The ISM-corrected intrinsic  $F$(Ly$\alpha$)/$F$(\ion{Mg}{2}) ratio for weakly active M dwarf exoplanet host stars is 10~$\pm$~3.

	\item The total FUV/NUV flux ratio is in the range~$\approx$~0.5~--~3 for all M dwarfs, including AD Leo. This ratio is 
 $\lesssim$~10$^{-2}$ for K1V dwarfs and $<$~10$^{-3}$ for the Sun.  The percentage of total UV energy from the star in the Ly$\alpha$ emission line is: 1)  37~--~75~\% for GJ 581, GJ 876, GJ 436, GJ 832, and AD Leo, 2) $<$ 7.4\% for GJ 1214, 3) $\approx$~0.7~\% for K1V dwarfs, and 4) $\approx$~0.04~\% for the Sun.  The FUV/NUV flux ratio can have a profound effect on the production of O$_{2}$ and O$_{3}$ in the atmospheres of Earth-like planets orbiting M dwarfs.  

	\item Fluorescent emission from hot H$_{2}$ ($T$(H$_{2}$)~$\approx$~2000~--~4000 K) is detected towards all four stars with moderate S/N COS spectra (GJ 581, GJ 876, GJ 436, and GJ 832).  A first order calculation suggests that this emission originates in the stellar photosphere, but this emission could also arise on the illuminated dayside of the planets in these systems or in a circumstellar envelope replenished by mass-loss from the planetary atmosphere.  

	\item Time-variability is detected in the chromospheric and transition region emission lines observed from all four stars with moderate S/N in the COS spectra (GJ 581, GJ 876, GJ 436, and GJ 832).  Variability with amplitudes 50~--~500\% is observed on time scales of 10$^{2}$~--~10$^{3}$ s.    This will most likely make UV transit studies of these systems challenging.  

\end{enumerate}

\acknowledgments
KF thanks Tom Ayres for technical assistance with STIS echelle spectra and enjoyable discussions about cool star atmospheres.  The quality and completeness of the manuscript was improved by the thoughtful comments of an anonymous referee.  The data presented here were obtained as part of the $HST$ Guest Observing program \#12464 and the COS Science Team Guaranteed Time programs \#12034 and \#12035.  
Brass Elephant.  
This work was supported by NASA grants HST-GO-12464.01, NNX08AC146, and NAS5-98043 to the University of Colorado at Boulder.  ~\nocite{mota05,yoshino96}


  \bibliography{ms}

\begin{thebibliography}{101}
\expandafter\ifx\csname natexlab\endcsname\relax\def\natexlab#1{#1}\fi

\bibitem[{{Anglada-Escud{\'e}} {et~al.}(2012){Anglada-Escud{\'e}}, {Arriagada},
  {Vogt}, {Rivera}, {Butler}, {Crane}, {Shectman}, {Thompson}, {Minniti},
  {Haghighipour}, {Carter}, {Tinney}, {Wittenmyer}, {Bailey}, {O'Toole},
  {Jones}, \& {Jenkins}}]{anglada_escude12}
{Anglada-Escud{\'e}}, G., {Arriagada}, P., {Vogt}, S.~S., {Rivera}, E.~J.,
  {Butler}, R.~P., {Crane}, J.~D., {Shectman}, S.~A., {Thompson}, I.~B.,
  {Minniti}, D., {Haghighipour}, N., {Carter}, B.~D., {Tinney}, C.~G.,
  {Wittenmyer}, R.~A., {Bailey}, J.~A., {O'Toole}, S.~J., {Jones}, H.~R.~A., \&
  {Jenkins}, J.~S. 2012, \apjl, 751, L16

\bibitem[{{Ayres}(1997)}]{ayres97}
{Ayres}, T.~R. 1997, \jgr, 102, 1641

\bibitem[{{Ayres}(2010)}]{ayres10}
---. 2010, \apjs, 187, 149

\bibitem[{{Bailey} {et~al.}(2009){Bailey}, {Butler}, {Tinney}, {Jones},
  {O'Toole}, {Carter}, \& {Marcy}}]{bailey09}
{Bailey}, J., {Butler}, R.~P., {Tinney}, C.~G., {Jones}, H.~R.~A., {O'Toole},
  S., {Carter}, B.~D., \& {Marcy}, G.~W. 2009, \apj, 690, 743

\bibitem[{{Bean} {et~al.}(2010){Bean}, {Miller-Ricci Kempton}, \&
  {Homeier}}]{bean10}
{Bean}, J.~L., {Miller-Ricci Kempton}, E., \& {Homeier}, D. 2010, \nat, 468,
  669

\bibitem[{{Ben-Jaffel}(2007)}]{jaffel07}
{Ben-Jaffel}, L. 2007, \apjl, 671, L61

\bibitem[{{Broadfoot} {et~al.}(1979){Broadfoot}, {Belton}, {Takacs}, {Sandel},
  {Shemansky}, {Holberg}, {Ajello}, {Moos}, {Atreya}, {Donahue}, {Bertaux},
  {Blamont}, {Strobel}, {McConnell}, {Goody}, {Dalgarno}, \&
  {McElroy}}]{broadfoot79}
{Broadfoot}, A.~L., {Belton}, M.~J., {Takacs}, P.~Z., {Sandel}, B.~R.,
  {Shemansky}, D.~E., {Holberg}, J.~B., {Ajello}, J.~M., {Moos}, H.~W.,
  {Atreya}, S.~K., {Donahue}, T.~M., {Bertaux}, J.~L., {Blamont}, J.~E.,
  {Strobel}, D.~F., {McConnell}, J.~C., {Goody}, R., {Dalgarno}, A., \&
  {McElroy}, M.~B. 1979, Science, 204, 979

\bibitem[{{Buccino} {et~al.}(2006){Buccino}, {Lemarchand}, \&
  {Mauas}}]{buccino06}
{Buccino}, A.~P., {Lemarchand}, G.~A., \& {Mauas}, P.~J.~D. 2006, \icarus, 183,
  491

\bibitem[{{Buccino} {et~al.}(2007){Buccino}, {Lemarchand}, \&
  {Mauas}}]{buccino07}
---. 2007, \icarus, 192, 582

\bibitem[{{Butler} {et~al.}(2004){Butler}, {Vogt}, {Marcy}, {Fischer},
  {Wright}, {Henry}, {Laughlin}, \& {Lissauer}}]{butler04}
{Butler}, R.~P., {Vogt}, S.~S., {Marcy}, G.~W., {Fischer}, D.~A., {Wright},
  J.~T., {Henry}, G.~W., {Laughlin}, G., \& {Lissauer}, J.~J. 2004, \apj, 617,
  580

\bibitem[{{Byrne} \& {Doyle}(1989)}]{byrne89}
{Byrne}, P.~B. \& {Doyle}, J.~G. 1989, \aap, 208, 159

\bibitem[{{Charbonneau} {et~al.}(2009){Charbonneau}, {Berta}, {Irwin}, {Burke},
  {Nutzman}, {Buchhave}, {Lovis}, {Bonfils}, {Latham}, {Udry}, {Murray-Clay},
  {Holman}, {Falco}, {Winn}, {Queloz}, {Pepe}, {Mayor}, {Delfosse}, \&
  {Forveille}}]{charbonneau09}
{Charbonneau}, D., {Berta}, Z.~K., {Irwin}, J., {Burke}, C.~J., {Nutzman}, P.,
  {Buchhave}, L.~A., {Lovis}, C., {Bonfils}, X., {Latham}, D.~W., {Udry}, S.,
  {Murray-Clay}, R.~A., {Holman}, M.~J., {Falco}, E.~E., {Winn}, J.~N.,
  {Queloz}, D., {Pepe}, F., {Mayor}, M., {Delfosse}, X., \& {Forveille}, T.
  2009, \nat, 462, 891

\bibitem[{{Correia} {et~al.}(2010){Correia}, {Couetdic}, {Laskar}, {Bonfils},
  {Mayor}, {Bertaux}, {Bouchy}, {Delfosse}, {Forveille}, {Lovis}, {Pepe},
  {Perrier}, {Queloz}, \& {Udry}}]{correia10}
{Correia}, A.~C.~M., {Couetdic}, J., {Laskar}, J., {Bonfils}, X., {Mayor}, M.,
  {Bertaux}, J.-L., {Bouchy}, F., {Delfosse}, X., {Forveille}, T., {Lovis}, C.,
  {Pepe}, F., {Perrier}, C., {Queloz}, D., \& {Udry}, S. 2010, \aap, 511, A21

\bibitem[{{Cravens}(1987)}]{cravens87}
{Cravens}, T.~E. 1987, \jgr, 92, 11083

\bibitem[{{Des Marais} {et~al.}(2002){Des Marais}, {Harwit}, {Jucks},
  {Kasting}, {Lin}, {Lunine}, {Schneider}, {Seager}, {Traub}, \&
  {Woolf}}]{marais02}
{Des Marais}, D.~J., {Harwit}, M.~O., {Jucks}, K.~W., {Kasting}, J.~F., {Lin},
  D.~N.~C., {Lunine}, J.~I., {Schneider}, J., {Seager}, S., {Traub}, W.~A., \&
  {Woolf}, N.~J. 2002, Astrobiology, 2, 153

\bibitem[{{D{\'e}sert} {et~al.}(2011){D{\'e}sert}, {Bean}, {Miller-Ricci
  Kempton}, {Berta}, {Charbonneau}, {Irwin}, {Fortney}, {Burke}, \&
  {Nutzman}}]{desert11}
{D{\'e}sert}, J.-M., {Bean}, J., {Miller-Ricci Kempton}, E., {Berta}, Z.~K.,
  {Charbonneau}, D., {Irwin}, J., {Fortney}, J., {Burke}, C.~J., \& {Nutzman},
  P. 2011, \apjl, 731, L40

\bibitem[{{Domagal-Goldman} {et~al.}(2012){Domagal-Goldman}, {Segura}, \&
  A.}]{goldman12}
{Domagal-Goldman}, S.~D., {Segura}, A., \& A., e. 2012, \apj, 0, 1

\bibitem[{{Ehrenreich} \& {D{\'e}sert}(2011)}]{ehrenreich11b}
{Ehrenreich}, D. \& {D{\'e}sert}, J.-M. 2011, \aap, 529, A136

\bibitem[{{Ehrenreich} {et~al.}(2011){Ehrenreich}, {Lecavelier Des Etangs}, \&
  {Delfosse}}]{ehrenreich11}
{Ehrenreich}, D., {Lecavelier Des Etangs}, A., \& {Delfosse}, X. 2011, \aap,
  529, A80

\bibitem[{{Feldman} {et~al.}(1993){Feldman}, {McGrath}, {Moos}, {Durrance},
  {Strobel}, \& {Davidsen}}]{feldman93}
{Feldman}, P.~D., {McGrath}, M.~A., {Moos}, H.~W., {Durrance}, S.~T.,
  {Strobel}, D.~F., \& {Davidsen}, A.~F. 1993, \apj, 406, 279

\bibitem[{{Fontenla} {et~al.}(2011){Fontenla}, {Harder}, {Livingston}, {Snow},
  \& {Woods}}]{fontenla11}
{Fontenla}, J.~M., {Harder}, J., {Livingston}, W., {Snow}, M., \& {Woods}, T.
  2011, Journal of Geophysical Research (Atmospheres), 116, 20108

\bibitem[{{Fossati} {et~al.}(2010){Fossati}, {Haswell}, {Froning}, {Hebb},
  {Holmes}, {Kolb}, {Helling}, {Carter}, {Wheatley}, {Collier Cameron},
  {Loeillet}, {Pollacco}, {Street}, {Stempels}, {Simpson}, {Udry}, {Joshi},
  {West}, {Skillen}, \& {Wilson}}]{fossati10}
{Fossati}, L., {Haswell}, C.~A., {Froning}, C.~S., {Hebb}, L., {Holmes}, S.,
  {Kolb}, U., {Helling}, C., {Carter}, A., {Wheatley}, P., {Collier Cameron},
  A., {Loeillet}, B., {Pollacco}, D., {Street}, R., {Stempels}, H.~C.,
  {Simpson}, E., {Udry}, S., {Joshi}, Y.~C., {West}, R.~G., {Skillen}, I., \&
  {Wilson}, D. 2010, \apjl, 714, L222

\bibitem[{{France} {et~al.}(2010{\natexlab{a}}){France}, {Linsky}, {Brown},
  {Froning}, \& {B{\'e}land}}]{france10b}
{France}, K., {Linsky}, J.~L., {Brown}, A., {Froning}, C.~S., \& {B{\'e}land},
  S. 2010{\natexlab{a}}, \apj, 715, 596

\bibitem[{{France} {et~al.}(2012{\natexlab{a}}){France}, {Linsky}, {Tian},
  {Froning}, \& {Roberge}}]{france12b}
{France}, K., {Linsky}, J.~L., {Tian}, F., {Froning}, C.~S., \& {Roberge}, A.
  2012{\natexlab{a}}, \apjl, 750, L32

\bibitem[{{France} {et~al.}(2007){France}, {Roberge}, {Lupu}, {Redfield}, \&
  {Feldman}}]{france07}
{France}, K., {Roberge}, A., {Lupu}, R.~E., {Redfield}, S., \& {Feldman}, P.~D.
  2007, \apj, 668, 1174

\bibitem[{{France} {et~al.}(2012{\natexlab{b}}){France}, {Schindhelm},
  {Herczeg}, {Brown}, {Abgrall}, {Alexander}, {Bergin}, {Brown}, {Linsky},
  {Roueff}, \& {Yang}}]{france12c}
{France}, K., {Schindhelm}, E., {Herczeg}, G.~J., {Brown}, A., {Abgrall}, H.,
  {Alexander}, R.~D., {Bergin}, E.~A., {Brown}, J.~M., {Linsky}, J.~L.,
  {Roueff}, E., \& {Yang}, H. 2012{\natexlab{b}}, ArXiv e-prints

\bibitem[{{France} {et~al.}(2010{\natexlab{b}}){France}, {Stocke}, {Yang},
  {Linsky}, {Wolven}, {Froning}, {Green}, \& {Osterman}}]{france10a}
{France}, K., {Stocke}, J.~T., {Yang}, H., {Linsky}, J.~L., {Wolven}, B.~C.,
  {Froning}, C.~S., {Green}, J.~C., \& {Osterman}, S.~N. 2010{\natexlab{b}},
  \apj, 712, 1277

\bibitem[{{Gizis} {et~al.}(2002){Gizis}, {Reid}, \& {Hawley}}]{gizis02}
{Gizis}, J.~E., {Reid}, I.~N., \& {Hawley}, S.~L. 2002, \aj, 123, 3356

\bibitem[{{Green} {et~al.}(2012){Green}, {Froning}, {Osterman}, {Ebbets},
  {Heap}, {Leitherer}, {Linsky}, {Savage}, {Sembach}, {Shull}, {Siegmund},
  {Snow}, {Spencer}, {Stern}, {Stocke}, {Welsh}, {B{\'e}land}, {Burgh},
  {Danforth}, {France}, {Keeney}, {McPhate}, {Penton}, {Andrews},
  {Brownsberger}, {Morse}, \& {Wilkinson}}]{green12}
{Green}, J.~C., {Froning}, C.~S., {Osterman}, S., {Ebbets}, D., {Heap}, S.~H.,
  {Leitherer}, C., {Linsky}, J.~L., {Savage}, B.~D., {Sembach}, K., {Shull},
  J.~M., {Siegmund}, O.~H.~W., {Snow}, T.~P., {Spencer}, J., {Stern}, S.~A.,
  {Stocke}, J., {Welsh}, B., {B{\'e}land}, S., {Burgh}, E.~B., {Danforth}, C.,
  {France}, K., {Keeney}, B., {McPhate}, J., {Penton}, S.~V., {Andrews}, J.,
  {Brownsberger}, K., {Morse}, J., \& {Wilkinson}, E. 2012, \apj, 744, 60

\bibitem[{{Gustin} {et~al.}(2010){Gustin}, {Stewart}, {G{\'e}rard}, \&
  {Esposito}}]{gustin10}
{Gustin}, J., {Stewart}, I., {G{\'e}rard}, J.-C., \& {Esposito}, L. 2010,
  \icarus, 210, 270

\bibitem[{{Haisch} {et~al.}(1991){Haisch}, {Strong}, \& {Rodono}}]{haisch91}
{Haisch}, B., {Strong}, K.~T., \& {Rodono}, M. 1991, \araa, 29, 275

\bibitem[{{Haswell} {et~al.}(2012){Haswell}, {Fosatti}, \& {et
  al.}}]{haswell12}
{Haswell}, C.~A., {Fosatti}, L., \& {et al.}, A. 2012, \apj, 0, 1

\bibitem[{{Hawley} {et~al.}(2003){Hawley}, {Allred}, {Johns-Krull}, {Fisher},
  {Abbett}, {Alekseev}, {Avgoloupis}, {Deustua}, {Gunn}, {Seiradakis}, {Sirk},
  \& {Valenti}}]{hawley03}
{Hawley}, S.~L., {Allred}, J.~C., {Johns-Krull}, C.~M., {Fisher}, G.~H.,
  {Abbett}, W.~P., {Alekseev}, I., {Avgoloupis}, S.~I., {Deustua}, S.~E.,
  {Gunn}, A., {Seiradakis}, J.~H., {Sirk}, M.~M., \& {Valenti}, J.~A. 2003,
  \apj, 597, 535

\bibitem[{{Hawley} {et~al.}(1996){Hawley}, {Gizis}, \& {Reid}}]{hawley96}
{Hawley}, S.~L., {Gizis}, J.~E., \& {Reid}, I.~N. 1996, \aj, 112, 2799

\bibitem[{{Hawley} \& {Pettersen}(1991)}]{hawley91}
{Hawley}, S.~L. \& {Pettersen}, B.~R. 1991, \apj, 378, 725

\bibitem[{{Herczeg} {et~al.}(2004){Herczeg}, {Wood}, {Linsky}, {Valenti}, \&
  {Johns-Krull}}]{herczeg04}
{Herczeg}, G.~J., {Wood}, B.~E., {Linsky}, J.~L., {Valenti}, J.~A., \&
  {Johns-Krull}, C.~M. 2004, \apj, 607, 369

\bibitem[{{Johnson} \& {Apps}(2009)}]{johnson09}
{Johnson}, J.~A. \& {Apps}, K. 2009, \apj, 699, 933

\bibitem[{{Jordan} {et~al.}(1977){Jordan}, {Brueckner}, {Bartoe}, {Sandlin}, \&
  {van Hoosier}}]{jordan77}
{Jordan}, C., {Brueckner}, G.~E., {Bartoe}, J.-D.~F., {Sandlin}, G.~D., \& {van
  Hoosier}, M.~E. 1977, \nat, 270, 326

\bibitem[{{Jura}(1975)}]{jura75a}
{Jura}, M. 1975, \apj, 197, 575

\bibitem[{{Kaltenegger} {et~al.}(2011){Kaltenegger}, {Segura}, \&
  {Mohanty}}]{kaltenegger11}
{Kaltenegger}, L., {Segura}, A., \& {Mohanty}, S. 2011, \apj, 733, 35

\bibitem[{{Kim} \& {Fox}(1994)}]{kim94}
{Kim}, Y.~H. \& {Fox}, J.~L. 1994, \icarus, 112, 310

\bibitem[{{Kowalski} {et~al.}(2009){Kowalski}, {Hawley}, {Hilton}, {Becker},
  {West}, {Bochanski}, \& {Sesar}}]{kowalski09}
{Kowalski}, A.~F., {Hawley}, S.~L., {Hilton}, E.~J., {Becker}, A.~C., {West},
  A.~A., {Bochanski}, J.~J., \& {Sesar}, B. 2009, \aj, 138, 633

\bibitem[{{Kriss}(2011)}]{kriss11}
{Kriss}, G.~A. 2011, {Improved Medium Resolution Line Spread Functions for COS
  FUV Spectra}, Tech. rep.

\bibitem[{{Kundurthy} {et~al.}(2011){Kundurthy}, {Agol}, {Becker}, {Barnes},
  {Williams}, \& {Mukadam}}]{kundurthy11}
{Kundurthy}, P., {Agol}, E., {Becker}, A.~C., {Barnes}, R., {Williams}, B., \&
  {Mukadam}, A. 2011, \apj, 731, 123

\bibitem[{{Lecavelier des Etangs} {et~al.}(2012){Lecavelier des Etangs},
  {Bourrier}, {Wheatley}, {Dupuy}, {Ehrenreich}, {Vidal-Madjar}, {H{\'e}brard},
  {Ballester}, {D{\'e}sert}, {Ferlet}, \& {Sing}}]{lecavelier12}
{Lecavelier des Etangs}, A., {Bourrier}, V., {Wheatley}, P.~J., {Dupuy}, H.,
  {Ehrenreich}, D., {Vidal-Madjar}, A., {H{\'e}brard}, G., {Ballester}, G.~E.,
  {D{\'e}sert}, J.-M., {Ferlet}, R., \& {Sing}, D.~K. 2012, \aap, 543, L4

\bibitem[{{Linsky} {et~al.}(2012{\natexlab{a}}){Linsky}, {France}, \&
  {Ayres}}]{linsky12c}
{Linsky}, J., {France}, K., \& {Ayres}, T. 2012{\natexlab{a}}, \apj, 1

\bibitem[{{Linsky} {et~al.}(2012{\natexlab{b}}){Linsky}, {Bushinsky}, {Ayres},
  {Fontenla}, \& {France}}]{linsky12a}
{Linsky}, J.~L., {Bushinsky}, R., {Ayres}, T., {Fontenla}, J., \& {France}, K.
  2012{\natexlab{b}}, \apj, 745, 25

\bibitem[{{Linsky} {et~al.}(1995){Linsky}, {Diplas}, {Wood}, {Brown}, {Ayres},
  \& {Savage}}]{linsky95}
{Linsky}, J.~L., {Diplas}, A., {Wood}, B.~E., {Brown}, A., {Ayres}, T.~R., \&
  {Savage}, B.~D. 1995, \apj, 451, 335

\bibitem[{{Linsky} {et~al.}(2006){Linsky}, {Draine}, {Moos}, {Jenkins}, {Wood},
  {Oliveira}, {Blair}, {Friedman}, {Gry}, {Knauth}, {Kruk}, {Lacour}, {Lehner},
  {Redfield}, {Shull}, {Sonneborn}, \& {Williger}}]{linsky06}
{Linsky}, J.~L., {Draine}, B.~T., {Moos}, H.~W., {Jenkins}, E.~B., {Wood},
  B.~E., {Oliveira}, C., {Blair}, W.~P., {Friedman}, S.~D., {Gry}, C.,
  {Knauth}, D., {Kruk}, J.~W., {Lacour}, S., {Lehner}, N., {Redfield}, S.,
  {Shull}, J.~M., {Sonneborn}, G., \& {Williger}, G.~M. 2006, \apj, 647, 1106

\bibitem[{{Linsky} \& {Wood}(1996)}]{linsky96}
{Linsky}, J.~L. \& {Wood}, B.~E. 1996, \apj, 463, 254

\bibitem[{{Linsky} {et~al.}(2010){Linsky}, {Yang}, {France}, {Froning},
  {Green}, {Stocke}, \& {Osterman}}]{linsky10}
{Linsky}, J.~L., {Yang}, H., {France}, K., {Froning}, C.~S., {Green}, J.~C.,
  {Stocke}, J.~T., \& {Osterman}, S.~N. 2010, \apj, 717, 1291

\bibitem[{{Markwardt}(2009)}]{markwardt09}
{Markwardt}, C.~B. 2009, in Astronomical Society of the Pacific Conference
  Series, Vol. 411, Astronomical Data Analysis Software and Systems XVIII, ed.
  D.~A. {Bohlender}, D.~{Durand}, \& P.~{Dowler}, 251

\bibitem[{{Mayor} {et~al.}(2009){Mayor}, {Bonfils}, {Forveille}, {Delfosse},
  {Udry}, {Bertaux}, {Beust}, {Bouchy}, {Lovis}, {Pepe}, {Perrier}, {Queloz},
  \& {Santos}}]{mayor09}
{Mayor}, M., {Bonfils}, X., {Forveille}, T., {Delfosse}, X., {Udry}, S.,
  {Bertaux}, J.-L., {Beust}, H., {Bouchy}, F., {Lovis}, C., {Pepe}, F.,
  {Perrier}, C., {Queloz}, D., \& {Santos}, N.~C. 2009, \aap, 507, 487

\bibitem[{{Mota} {et~al.}(2005){Mota}, {Parafita}, {Giuliani},
  {Hubin-Franskin}, {Lourenco}, {Garcia}, {Hoffmann}, {Mason}, {Ribeiro},
  {Raposo}, \& {Limao-Vieira}}]{mota05}
{Mota}, R., {Parafita}, R., {Giuliani}, A., {Hubin-Franskin}, M., {Lourenco},
  J.~M.~C., {Garcia}, G., {Hoffmann}, S.~V., {Mason}, N.~J., {Ribeiro}, P.~A.,
  {Raposo}, M., \& {Limao-Vieira}, P. 2005, Chemical Physics Letters, 416, 152

\bibitem[{{Murray-Clay} {et~al.}(2009){Murray-Clay}, {Chiang}, \&
  {Murray}}]{murray-clay09}
{Murray-Clay}, R.~A., {Chiang}, E.~I., \& {Murray}, N. 2009, \apj, 693, 23

\bibitem[{{Osterman} {et~al.}(2011){Osterman}, {Green}, {Froning},
  {B{\'e}land}, {Burgh}, {France}, {Penton}, {Delker}, {Ebbets}, {Sahnow},
  {Bacinski}, {Kimble}, {Andrews}, {Wilkinson}, {McPhate}, {Siegmund}, {Ake},
  {Aloisi}, {Biagetti}, {Diaz}, {Dixon}, {Friedman}, {Ghavamian}, {Goudfrooij},
  {Hartig}, {Keyes}, {Lennon}, {Massa}, {Niemi}, {Oliveira}, {Osten},
  {Proffitt}, {Smith}, \& {Soderblom}}]{osterman11}
{Osterman}, S., {Green}, J., {Froning}, C., {B{\'e}land}, S., {Burgh}, E.,
  {France}, K., {Penton}, S., {Delker}, T., {Ebbets}, D., {Sahnow}, D.,
  {Bacinski}, J., {Kimble}, R., {Andrews}, J., {Wilkinson}, E., {McPhate}, J.,
  {Siegmund}, O., {Ake}, T., {Aloisi}, A., {Biagetti}, C., {Diaz}, R., {Dixon},
  W., {Friedman}, S., {Ghavamian}, P., {Goudfrooij}, P., {Hartig}, G., {Keyes},
  C., {Lennon}, D., {Massa}, D., {Niemi}, S., {Oliveira}, C., {Osten}, R.,
  {Proffitt}, C., {Smith}, T., \& {Soderblom}, D. 2011, \apss, 157

\bibitem[{{Perrin} {et~al.}(1988){Perrin}, {Cayrel de Strobel}, \&
  {Dennefeld}}]{perrin88}
{Perrin}, M.-N., {Cayrel de Strobel}, G., \& {Dennefeld}, M. 1988, \aap, 191,
  237

\bibitem[{{Pickles}(1998)}]{pickles98}
{Pickles}, A.~J. 1998, \pasp, 110, 863

\bibitem[{{Pont} {et~al.}(2009){Pont}, {Gilliland}, {Knutson}, {Holman}, \&
  {Charbonneau}}]{pont09}
{Pont}, F., {Gilliland}, R.~L., {Knutson}, H., {Holman}, M., \& {Charbonneau},
  D. 2009, \mnras, 393, L6

\bibitem[{{Poppenhaeger} {et~al.}(2010){Poppenhaeger}, {Robrade}, \&
  {Schmitt}}]{poppenhaeger10}
{Poppenhaeger}, K., {Robrade}, J., \& {Schmitt}, J.~H.~M.~M. 2010, \aap, 515,
  A98+

\bibitem[{{Rauer} {et~al.}(2011){Rauer}, {Gebauer}, {Paris}, {Cabrera},
  {Godolt}, {Grenfell}, {Belu}, {Selsis}, {Hedelt}, \& {Schreier}}]{rauer11}
{Rauer}, H., {Gebauer}, S., {Paris}, P.~V., {Cabrera}, J., {Godolt}, M.,
  {Grenfell}, J.~L., {Belu}, A., {Selsis}, F., {Hedelt}, P., \& {Schreier}, F.
  2011, \aap, 529, A8

\bibitem[{{Rauscher} \& {Marcy}(2006)}]{rauscher06}
{Rauscher}, E. \& {Marcy}, G.~W. 2006, \pasp, 118, 617

\bibitem[{{Redfield} {et~al.}(2002){Redfield}, {Linsky}, {Ake}, {Ayres},
  {Dupree}, {Robinson}, {Wood}, \& {Young}}]{redfield02}
{Redfield}, S., {Linsky}, J.~L., {Ake}, T.~B., {Ayres}, T.~R., {Dupree}, A.~K.,
  {Robinson}, R.~D., {Wood}, B.~E., \& {Young}, P.~R. 2002, \apj, 581, 626

\bibitem[{{Ribas} {et~al.}(2005){Ribas}, {Guinan}, {G{\"u}del}, \&
  {Audard}}]{ribas05}
{Ribas}, I., {Guinan}, E.~F., {G{\"u}del}, M., \& {Audard}, M. 2005, \apj, 622,
  680

\bibitem[{{Rivera} {et~al.}(2010){Rivera}, {Laughlin}, {Butler}, {Vogt},
  {Haghighipour}, \& {Meschiari}}]{rivera10}
{Rivera}, E.~J., {Laughlin}, G., {Butler}, R.~P., {Vogt}, S.~S.,
  {Haghighipour}, N., \& {Meschiari}, S. 2010, \apj, 719, 890

\bibitem[{{Rivera} {et~al.}(2005){Rivera}, {Lissauer}, {Butler}, {Marcy},
  {Vogt}, {Fischer}, {Brown}, {Laughlin}, \& {Henry}}]{rivera05}
{Rivera}, E.~J., {Lissauer}, J.~J., {Butler}, R.~P., {Marcy}, G.~W., {Vogt},
  S.~S., {Fischer}, D.~A., {Brown}, T.~M., {Laughlin}, G., \& {Henry}, G.~W.
  2005, \apj, 634, 625

\bibitem[{{Rojas-Ayala} {et~al.}(2010){Rojas-Ayala}, {Covey}, {Muirhead}, \&
  {Lloyd}}]{rojas10}
{Rojas-Ayala}, B., {Covey}, K.~R., {Muirhead}, P.~S., \& {Lloyd}, J.~P. 2010,
  \apjl, 720, L113

\bibitem[{{Sanz-Forcada} {et~al.}(2011){Sanz-Forcada}, {Micela}, {Ribas},
  {Pollock}, {Eiroa}, {Velasco}, {Solano}, \&
  {Garc{\'{\i}}a-{\'A}lvarez}}]{forcada11}
{Sanz-Forcada}, J., {Micela}, G., {Ribas}, I., {Pollock}, A.~M.~T., {Eiroa},
  C., {Velasco}, A., {Solano}, E., \& {Garc{\'{\i}}a-{\'A}lvarez}, D. 2011,
  \aap, 532, A6

\bibitem[{{Seager} {et~al.}(2009){Seager}, {Deming}, \& {Valenti}}]{seager09}
{Seager}, S., {Deming}, D., \& {Valenti}, J.~A. 2009, {Transiting Exoplanets
  with JWST}, ed. H.~A. {Thronson}, M.~{Stiavelli}, \& A.~{Tielens}, 123

\bibitem[{{Segura} {et~al.}(2005){Segura}, {Kasting}, {Meadows}, {Cohen},
  {Scalo}, {Crisp}, {Butler}, \& {Tinetti}}]{segura05}
{Segura}, A., {Kasting}, J.~F., {Meadows}, V., {Cohen}, M., {Scalo}, J.,
  {Crisp}, D., {Butler}, R.~A.~H., \& {Tinetti}, G. 2005, Astrobiology, 5, 706

\bibitem[{{Segura} {et~al.}(2010){Segura}, {Walkowicz}, {Meadows}, {Kasting},
  \& {Hawley}}]{segura10}
{Segura}, A., {Walkowicz}, L.~M., {Meadows}, V., {Kasting}, J., \& {Hawley}, S.
  2010, Astrobiology, 10, 751

\bibitem[{{Selsis} {et~al.}(2007){Selsis}, {Kasting}, {Levrard}, {Paillet},
  {Ribas}, \& {Delfosse}}]{selsis07}
{Selsis}, F., {Kasting}, J.~F., {Levrard}, B., {Paillet}, J., {Ribas}, I., \&
  {Delfosse}, X. 2007, \aap, 476, 1373

\bibitem[{{Shull}(1978)}]{shull78}
{Shull}, J.~M. 1978, \apj, 224, 841

\bibitem[{{Stevenson} {et~al.}(2012){Stevenson}, {Harrington}, {Lust}, {Lewis},
  {Montagnier}, {Moses}, {Visscher}, {Blecic}, {Hardy}, {Cubillos}, \&
  {Campo}}]{stevenson12}
{Stevenson}, K.~B., {Harrington}, J., {Lust}, N.~B., {Lewis}, N.~K.,
  {Montagnier}, G., {Moses}, J.~I., {Visscher}, C., {Blecic}, J., {Hardy},
  R.~A., {Cubillos}, P., \& {Campo}, C.~J. 2012, \apj, 755, 9

\bibitem[{{Tian} {et~al.}(2012){Tian}, {France}, \& A.}]{tian12}
{Tian}, F., {France}, F., \& A., e. 2012, \apj, 0, 1

\bibitem[{{Tian} {et~al.}(2008){Tian}, {Kasting}, {Liu}, \& {Roble}}]{tian08}
{Tian}, F., {Kasting}, J.~F., {Liu}, H.-L., \& {Roble}, R.~G. 2008, Journal of
  Geophysical Research (Planets), 113, 5008

\bibitem[{{Torres}(2007)}]{torres07}
{Torres}, G. 2007, \apjl, 671, L65

\bibitem[{{Tremblin} \& {Chiang}(2012)}]{tremblin12}
{Tremblin}, P. \& {Chiang}, E. 2012, ArXiv e-prints

\bibitem[{{Tuomi}(2011)}]{tuomi11}
{Tuomi}, M. 2011, \aap, 528, L5

\bibitem[{{Vidal-Madjar} {et~al.}(2004){Vidal-Madjar}, {D{\'e}sert},
  {Lecavelier des Etangs}, {H{\'e}brard}, {Ballester}, {Ehrenreich}, {Ferlet},
  {McConnell}, {Mayor}, \& {Parkinson}}]{vidal04}
{Vidal-Madjar}, A., {D{\'e}sert}, J.-M., {Lecavelier des Etangs}, A.,
  {H{\'e}brard}, G., {Ballester}, G.~E., {Ehrenreich}, D., {Ferlet}, R.,
  {McConnell}, J.~C., {Mayor}, M., \& {Parkinson}, C.~D. 2004, \apjl, 604, L69

\bibitem[{{Vidal-Madjar} {et~al.}(2003){Vidal-Madjar}, {Lecavelier des Etangs},
  {D{\'e}sert}, {Ballester}, {Ferlet}, {H{\'e}brard}, \& {Mayor}}]{vidal03}
{Vidal-Madjar}, A., {Lecavelier des Etangs}, A., {D{\'e}sert}, J.-M.,
  {Ballester}, G.~E., {Ferlet}, R., {H{\'e}brard}, G., \& {Mayor}, M. 2003,
  \nat, 422, 143

\bibitem[{{Vidal-Madjar} {et~al.}(2008){Vidal-Madjar}, {Lecavelier des Etangs},
  {D{\'e}sert}, {Ballester}, {Ferlet}, {H{\'e}brard}, \& {Mayor}}]{vidal08}
---. 2008, \apjl, 676, L57

\bibitem[{{von Braun} {et~al.}(2012){von Braun}, {Boyajian}, {Kane}, {Hebb},
  {van Belle}, {Farrington}, {Ciardi}, {Knutson}, {ten Brummelaar},
  {L{\'o}pez-Morales}, {McAlister}, {Schaefer}, {Ridgway}, {Collier Cameron},
  {Goldfinger}, {Turner}, {Sturmann}, \& {Sturmann}}]{vonbraun12}
{von Braun}, K., {Boyajian}, T.~S., {Kane}, S.~R., {Hebb}, L., {van Belle},
  G.~T., {Farrington}, C., {Ciardi}, D.~R., {Knutson}, H.~A., {ten Brummelaar},
  T.~A., {L{\'o}pez-Morales}, M., {McAlister}, H.~A., {Schaefer}, G.,
  {Ridgway}, S., {Collier Cameron}, A., {Goldfinger}, P.~J., {Turner}, N.~H.,
  {Sturmann}, L., \& {Sturmann}, J. 2012, \apj, 753, 171

\bibitem[{{von Braun} {et~al.}(2011){von Braun}, {Boyajian}, {Kane}, {van
  Belle}, {Ciardi}, {L{\'o}pez-Morales}, {McAlister}, {Henry}, {Jao}, {Riedel},
  {Subasavage}, {Schaefer}, {ten Brummelaar}, {Ridgway}, {Sturmann},
  {Sturmann}, {Mazingue}, {Turner}, {Farrington}, {Goldfinger}, \&
  {Boden}}]{vonbraun11}
{von Braun}, K., {Boyajian}, T.~S., {Kane}, S.~R., {van Belle}, G.~T.,
  {Ciardi}, D.~R., {L{\'o}pez-Morales}, M., {McAlister}, H.~A., {Henry}, T.~J.,
  {Jao}, W.-C., {Riedel}, A.~R., {Subasavage}, J.~P., {Schaefer}, G., {ten
  Brummelaar}, T.~A., {Ridgway}, S., {Sturmann}, L., {Sturmann}, J.,
  {Mazingue}, J., {Turner}, N.~H., {Farrington}, C., {Goldfinger}, P.~J., \&
  {Boden}, A.~F. 2011, \apjl, 729, L26+

\bibitem[{{Walkowicz} {et~al.}(2011){Walkowicz}, {Basri}, {Batalha},
  {Gilliland}, {Jenkins}, {Borucki}, {Koch}, {Caldwell}, {Dupree}, {Latham},
  {Meibom}, {Howell}, {Brown}, \& {Bryson}}]{walkowicz11}
{Walkowicz}, L.~M., {Basri}, G., {Batalha}, N., {Gilliland}, R.~L., {Jenkins},
  J., {Borucki}, W.~J., {Koch}, D., {Caldwell}, D., {Dupree}, A.~K., {Latham},
  D.~W., {Meibom}, S., {Howell}, S., {Brown}, T.~M., \& {Bryson}, S. 2011, \aj,
  141, 50

\bibitem[{{Walkowicz} \& {Hawley}(2009)}]{walkowicz09}
{Walkowicz}, L.~M. \& {Hawley}, S.~L. 2009, \aj, 137, 3297

\bibitem[{{Walkowicz} {et~al.}(2008){Walkowicz}, {Johns-Krull}, \&
  {Hawley}}]{walkowicz08}
{Walkowicz}, L.~M., {Johns-Krull}, C.~M., \& {Hawley}, S.~L. 2008, \apj, 677,
  593

\bibitem[{{Welsh} {et~al.}(2007){Welsh}, {Wheatley}, {Seibert}, {Browne},
  {West}, {Siegmund}, {Barlow}, {Forster}, {Friedman}, {Martin}, {Morrissey},
  {Small}, {Wyder}, {Schiminovich}, {Neff}, \& {Rich}}]{welsh07}
{Welsh}, B.~Y., {Wheatley}, J.~M., {Seibert}, M., {Browne}, S.~E., {West},
  A.~A., {Siegmund}, O.~H.~W., {Barlow}, T.~A., {Forster}, K., {Friedman},
  P.~G., {Martin}, D.~C., {Morrissey}, P., {Small}, T., {Wyder}, T.,
  {Schiminovich}, D., {Neff}, S., \& {Rich}, R.~M. 2007, \apjs, 173, 673

\bibitem[{{West} {et~al.}(2004){West}, {Hawley}, {Walkowicz}, {Covey},
  {Silvestri}, {Raymond}, {Harris}, {Munn}, {McGehee}, {Ivezi{\'c}}, \&
  {Brinkmann}}]{west04}
{West}, A.~A., {Hawley}, S.~L., {Walkowicz}, L.~M., {Covey}, K.~R.,
  {Silvestri}, N.~M., {Raymond}, S.~N., {Harris}, H.~C., {Munn}, J.~A.,
  {McGehee}, P.~M., {Ivezi{\'c}}, {\v Z}., \& {Brinkmann}, J. 2004, \aj, 128,
  426

\bibitem[{{Wolven} \& {Feldman}(1998)}]{wolven98}
{Wolven}, B.~C. \& {Feldman}, P.~D. 1998, \grl, 25, 1537

\bibitem[{{Wolven} {et~al.}(1997){Wolven}, {Feldman}, {Strobel}, \&
  {McGrath}}]{wolven97}
{Wolven}, B.~C., {Feldman}, P.~D., {Strobel}, D.~F., \& {McGrath}, M.~A. 1997,
  \apj, 475, 835

\bibitem[{{Wood} {et~al.}(2000){Wood}, {Ambruster}, {Brown}, \&
  {Linsky}}]{wood00}
{Wood}, B.~E., {Ambruster}, C.~W., {Brown}, A., \& {Linsky}, J.~L. 2000, \apj,
  542, 411

\bibitem[{{Wood} \& {Linsky}(1998)}]{wood98}
{Wood}, B.~E. \& {Linsky}, J.~L. 1998, \apj, 492, 788

\bibitem[{{Wood} {et~al.}(1997){Wood}, {Linsky}, \& {Ayres}}]{wood97}
{Wood}, B.~E., {Linsky}, J.~L., \& {Ayres}, T.~R. 1997, \apj, 478, 745

\bibitem[{{Wood} {et~al.}(2005){Wood}, {Redfield}, {Linsky}, {M{\"u}ller}, \&
  {Zank}}]{wood05}
{Wood}, B.~E., {Redfield}, S., {Linsky}, J.~L., {M{\"u}ller}, H.-R., \& {Zank},
  G.~P. 2005, \apjs, 159, 118

\bibitem[{{Woods} {et~al.}(2009){Woods}, {Chamberlin}, {Harder}, {Hock},
  {Snow}, {Eparvier}, {Fontenla}, {McClintock}, \& {Richard}}]{woods09}
{Woods}, T.~N., {Chamberlin}, P.~C., {Harder}, J.~W., {Hock}, R.~A., {Snow},
  M., {Eparvier}, F.~G., {Fontenla}, J., {McClintock}, W.~E., \& {Richard},
  E.~C. 2009, \grl, 36, 1101

\bibitem[{{Wordsworth} {et~al.}(2010){Wordsworth}, {Forget}, {Selsis},
  {Madeleine}, {Millour}, \& {Eymet}}]{wordsworth10}
{Wordsworth}, R.~D., {Forget}, F., {Selsis}, F., {Madeleine}, J.-B., {Millour},
  E., \& {Eymet}, V. 2010, \aap, 522, A22+

\bibitem[{{Wordsworth} {et~al.}(2011){Wordsworth}, {Forget}, {Selsis},
  {Millour}, {Charnay}, \& {Madeleine}}]{wordsworth11}
{Wordsworth}, R.~D., {Forget}, F., {Selsis}, F., {Millour}, E., {Charnay}, B.,
  \& {Madeleine}, J.-B. 2011, \apjl, 733, L48+

\bibitem[{{Yelle}(2004)}]{yelle04}
{Yelle}, R.~V. 2004, \icarus, 170, 167

\bibitem[{{Yi} {et~al.}(2007){Yi}, {Park}, \& {Lee}}]{yi07}
{Yi}, W., {Park}, J., \& {Lee}, J. 2007, Chemical Physics Letters, 439, 46

\bibitem[{{Yoshino} {et~al.}(1996){Yoshino}, {Esmond}, {Sun}, {Parkinson},
  {Ito}, \& {Matsui}}]{yoshino96}
{Yoshino}, K., {Esmond}, J.~R., {Sun}, Y., {Parkinson}, W.~H., {Ito}, K., \&
  {Matsui}, T. 1996, \jqsrt, 55, 53

\end{thebibliography}


\begin{deluxetable*}{lc|ccc}
\tabletypesize{\normalsize}
\tablecaption{Brightest Chromospheric and Transition Region M dwarf Emission Line Fluxes (ergs cm$^{-2}$ s$^{-1}$)\tablenotemark{a}. \label{lya_lines}}
\tablewidth{0pt}
\tablehead{
\colhead{Target} & \colhead{$d$ (pc)} & \colhead{$F$(Ly$\alpha$)\tablenotemark{b}}   &  \colhead{$F$(\ion{C}{4})\tablenotemark{c}} & \colhead{$F$(\ion{Mg}{2})\tablenotemark{c}}    
}
\startdata
GJ 581     &   6.3   &   3.0~$\times$~10$^{-13}$     &    1.80~($\pm$~0.16)~$\times$~10$^{-15}$    &    2.13~($\pm$~0.13)~$\times$~10$^{-14}$ \\
GJ 876     &   4.7   &   4.4~$\times$~10$^{-13}$    &    1.99~($\pm$~0.17)~$\times$~10$^{-14}$    &    3.37~($\pm$~0.23)~$\times$~10$^{-14}$ \\
GJ 436     &   10.2   &   3.5~$\times$~10$^{-13}$    &    1.38~($\pm$~0.11)~$\times$~10$^{-15}$    &    2.30~($\pm$~0.13)~$\times$~10$^{-14}$ \\
GJ 832     &   4.9   &   5.0~$\times$~10$^{-12}$     &    7.30~($\pm$~0.48)~$\times$~10$^{-15}$    &    2.98~($\pm$~0.18)~$\times$~10$^{-13}$ \\
GJ 667C    &   6.9    &   7.6~$\times$~10$^{-13}$   &   $\cdots$    &   5.59~($\pm$~0.47)~$\times$~10$^{-14}$   \\
GJ 1214    &   13.0   &   $<$ 2.4~$\times$~10$^{-15}$     &    2.62~($\pm$~0.47)~$\times$~10$^{-16}$    &    2.21~($\pm$~0.15)~$\times$~10$^{-15}$ 
\enddata
\tablenotetext{a}{Flux measurements are averaged over all exposure times. } 
\tablenotetext{b}{Uncertainty on the reconstructed Ly$\alpha$ flux is estimated to be between 10~--~20\% for the stars observed with E140M (GJ 832 and GJ 667C) and 15~--~30\% for stars observed with G140M (GJ 581, GJ 876, and GJ 436).} 
\tablenotetext{c}{\ion{C}{4} and \ion{Mg}{2} fluxes are Gaussian fits to both lines of the doublets.  } 
\end{deluxetable*}

\end{document}